\documentclass[epj,nopacs]{svjour}
\pdfoutput=1 
\usepackage{graphicx}
\def\ttbar {\mbox {$t\bar{t}$}}

\def\ifb {\mbox {$~{\rm fb}^{-1}$}}

\def\lumierr{\mbox{$20.3 \pm 0.6 \ifb$}}

\def\pt{\mbox{$p_{\text{T}}$}}

\def\MET{\ensuremath{E_{\mathrm{T}}^{\mathrm{miss}}}}
\def\kt{\mbox{$k_t$}}

\def\MTW{\mbox{$m_\mathrm{T}(\ell , \MET)$}}
\def\mfmet{\ensuremath{m(f_{\mathrm{met}})}}
\def\mvmet{\ensuremath{m(v_{\mathrm{met}})}}
\def\vmet{\ensuremath{v_{\mathrm{met}}}}
\def\fmet{\ensuremath{f_{\mathrm{met}}}}
\def\ares{\ensuremath{a_{\mathrm{res}}}}
\def\anonres{\ensuremath{a_{\mathrm{non\mbox{-}res}}}}
\def\deltaphilb{\ensuremath{\Delta\phi(\ell,b)}}

\DeclareGraphicsExtensions{.pdf,.png}

\RequirePackage[numbers,sort&compress]{natbib}

\newcommand{\rhtitle}{
Search for invisible particles produced in association with single-top-quarks in proton-proton collisions at $\sqrt{s}=\mathrm{8~TeV}$ with the ATLAS detector
}
\newcommand{\rhabstract}{
A search for the production of single-top-quarks in association with missing energy is performed in 
proton--proton collisions at a centre-of-mass energy of $\sqrt{s}=\mathrm{8~TeV}$ with the ATLAS experiment at the Large Hadron Collider using
data collected in 2012, corresponding to an integrated luminosity of $20.3$~fb$^{-1}$.
In this search, the $W$ boson from the top quark is required to decay into an electron or a muon and a neutrino.
No deviation from the Standard Model prediction is observed, and upper limits are set on the production cross-section
for resonant and non-resonant production of an invisible exotic state in association with a right-handed top quark.
In the case of resonant production, for a spin-$0$ resonance with a mass of $500$~GeV, an effective coupling strength above $0.15$ 
is excluded  at 95$\%$ confidence level for the top quark and an invisible spin-$1/2$ state with mass between $0$~GeV and $100$~GeV. 
In the case of non-resonant production, an effective coupling strength above $0.2$ is excluded  at 95$\%$ confidence level for the top quark and an invisible spin-$1$ 
state with mass between $0$~GeV and $657$~GeV.
\\
{\bf Keywords } ATLAS - LHC - proton--proton collisions - top quark - single-top-quark - monotop
}

\usepackage{preprintcover}  
\PreprintCoverPaperTitle{\rhtitle}  
\PreprintIdNumber{CERN-PH-EP-2014-231}  
\PreprintCoverAbstract{\rhabstract}  
\PreprintJournalName{Eur.~Phys.~J.~C}  

\usepackage[hyperindex,breaklinks]{hyperref}

\usepackage{atlasphysics}
\usepackage{subfigure}
\usepackage{mathrsfs}
\usepackage{multirow}
\usepackage{extra_definitions}
\usepackage{url}
\usepackage{verbatim}
\usepackage{rotating}

\begin{document}
\title{\rhtitle}
\author{The ATLAS Collaboration} 

\institute{CERN, CH-1211 Geneva 23. Switzerland}

\date{Received: date / Revised version: date}

\abstract{
\rhabstract
\PACS{
      {PACS-key}{discribing text of that key}   \and
      {PACS-key}{discribing text of that key}
     } 
} 
%
\titlerunning{Search for invisible particles produced in association with single-top-quarks}
\maketitle
\section{Introduction}
\label{sec:into}
Many theories beyond the Standard Model (BSM) predict enhanced production of events
with large missing energy in association with a single reconstructed object.
Such events have been searched for at the Large Hadron Collider (LHC), when the single object is either a photon~\cite{MonoPhoton,MonoPhotonCMS},
a jet~\cite{monojet,monojetCMS}, or a $W$ or $Z$ boson~\cite{monoWZ,monoZ}.

This paper presents a search for singly produced top quarks in association with significant missing energy,
corresponding to the associated production of one or several undetected neutral particles, and without any other reconstructed object.
These neutral particles can be either stable and/or weakly interacting with ordinary matter -- providing an interesting interpretation
in terms of dark-matter candidates -- or long-lived and decaying outside of the detector.
The observation of such final states, commonly referred to as monotop events, would be evidence for new phenomena.
Moreover, processes involving top quarks are sensitive to BSM physics, due to the large mass of this Standard Model (SM) particle
which is close to the electroweak symmetry-breaking scale.

No such process is possible in the SM at tree level: the direct production of a top quark and a $Z$ boson
decaying into a pair of neutrinos, without any additional quark, is suppressed by the Glashow--Iliopoulos--Maiani mechanism~\cite{GIM}.

This search is performed with the ATLAS detector~\cite{ATLASPaper} in $pp$ collisions at $\mathrm{\sqrt{s}=8~TeV}$
with the data collected in 2012 at the LHC and corresponding to an integrated luminosity of $\mathrm{20.3~fb^{-1}}$.
The ATLAS detector covers the pseudorapidity
range $|\eta| <$ 4.9 and the full azimuthal angle $\phi$.\footnote{The ATLAS experiment uses a right-handed
coordinate system with its origin at the nominal interaction point (IP)
in the centre of the detector and the $z$-axis along the beam pipe.
The $x$-axis points from the IP to the centre of the LHC ring, and the $y$-axis points upwards.
Cylindrical coordinates $(r,\phi)$ are used in the transverse plane, $\phi$ being the azimuthal angle around the beam pipe.
The pseudorapidity is defined in terms of the polar angle $\theta$ as $\eta=-\ln\tan(\theta/2)$.}
It consists of an inner tracking detector covering
the pseudorapidity range $|\eta| <$ 2.5 surrounded by
a superconducting solenoid, electromagnetic and
hadronic calorimeters, and an external muon spectrometer
with large superconducting toroidal magnets.

The search is based on the analysis of top-quark events where the $W$ boson from the top quark decays into a lepton and a neutrino.
Previous results of a search for monotop production, exploiting the case of fully hadronic top-quark decays,
have been published by the CDF Collaboration using $p\overline{p}$ collision data at $\mathrm{\sqrt{s}=1.96~TeV}$,
corresponding to an integrated luminosity of $\mathrm{7.7~fb^{-1}}$~\cite{CDFmonotop},
and more recently by the CMS Collaboration using $pp$ collision data at $\mathrm{\sqrt{s}=8~TeV}$,
corresponding to an integrated luminosity of $\mathrm{19.7~fb^{-1}}$~\cite{CMSmonotop}.

\section{Signal models}
\label{sec:signal}

Many theoretical models predicting the production of monotop events in hadron colliders have been proposed.
In a first class of theories, a charged resonance is produced by down-type antiquark fusion
and decays into a top quark and a neutral particle, as in $SU(5)$ models~\cite{SU5}, in R-parity-violating SUSY~\cite{RPVSUSY}
or in hylogenesis models~\cite{Hylogenesis1,Hylogenesis2}.
In a second class of theories, the monotop final state is produced through a non-resonant process,
as in R-parity-conserving SUSY~\cite{RPCSUSY}, or in models where an interaction of a gluon with an up-type quark
allows production of a set of invisible particles via a $u$--$t$ or $c$--$t$ coupling~\cite{Aguilar99,Kamenik,EAlvarezetal,Kumaretal}.

\begin{figure*}[!h!tb]
\centering
\includegraphics[width=0.32\textwidth]{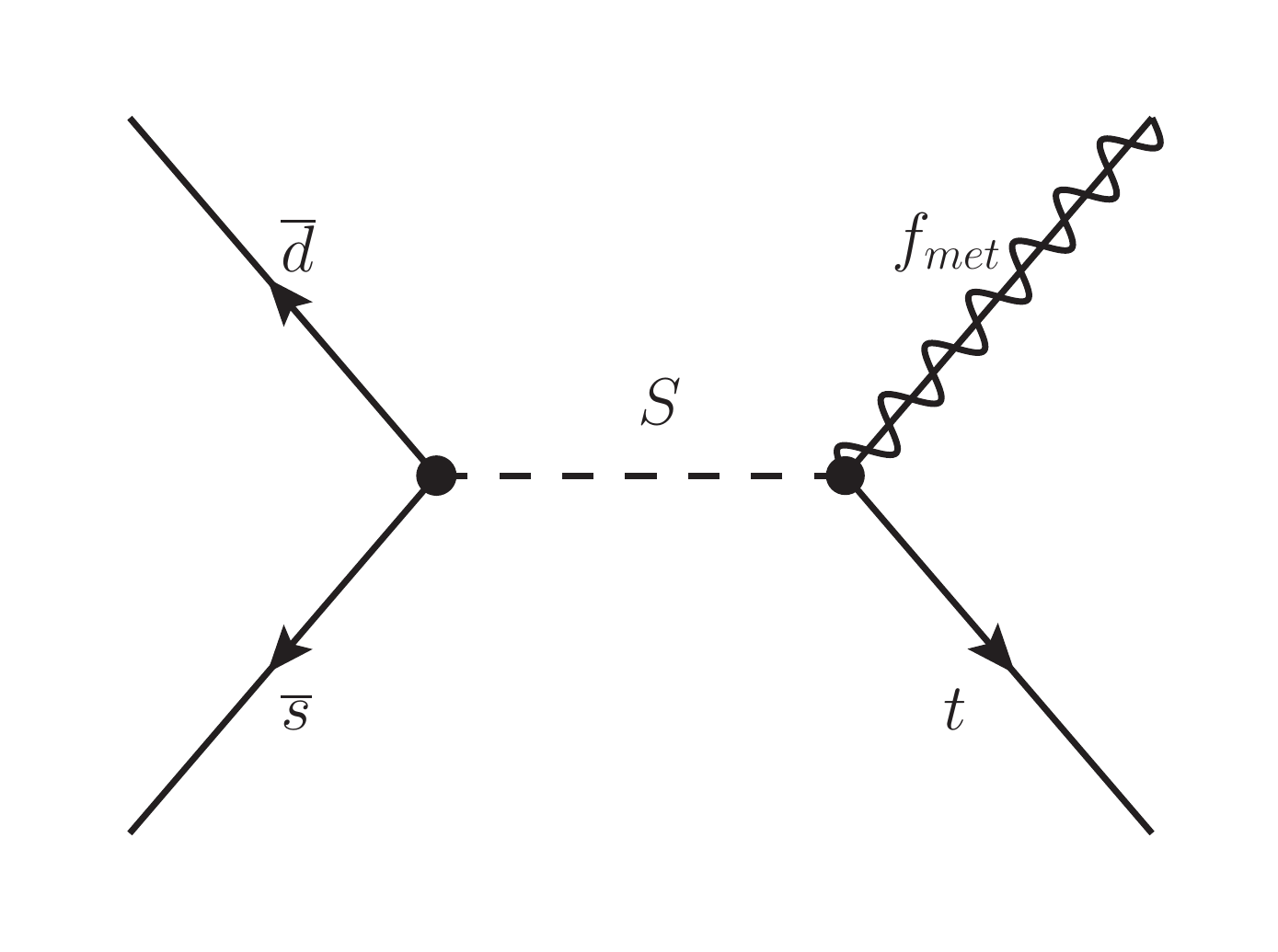}
\includegraphics[width=0.32\textwidth]{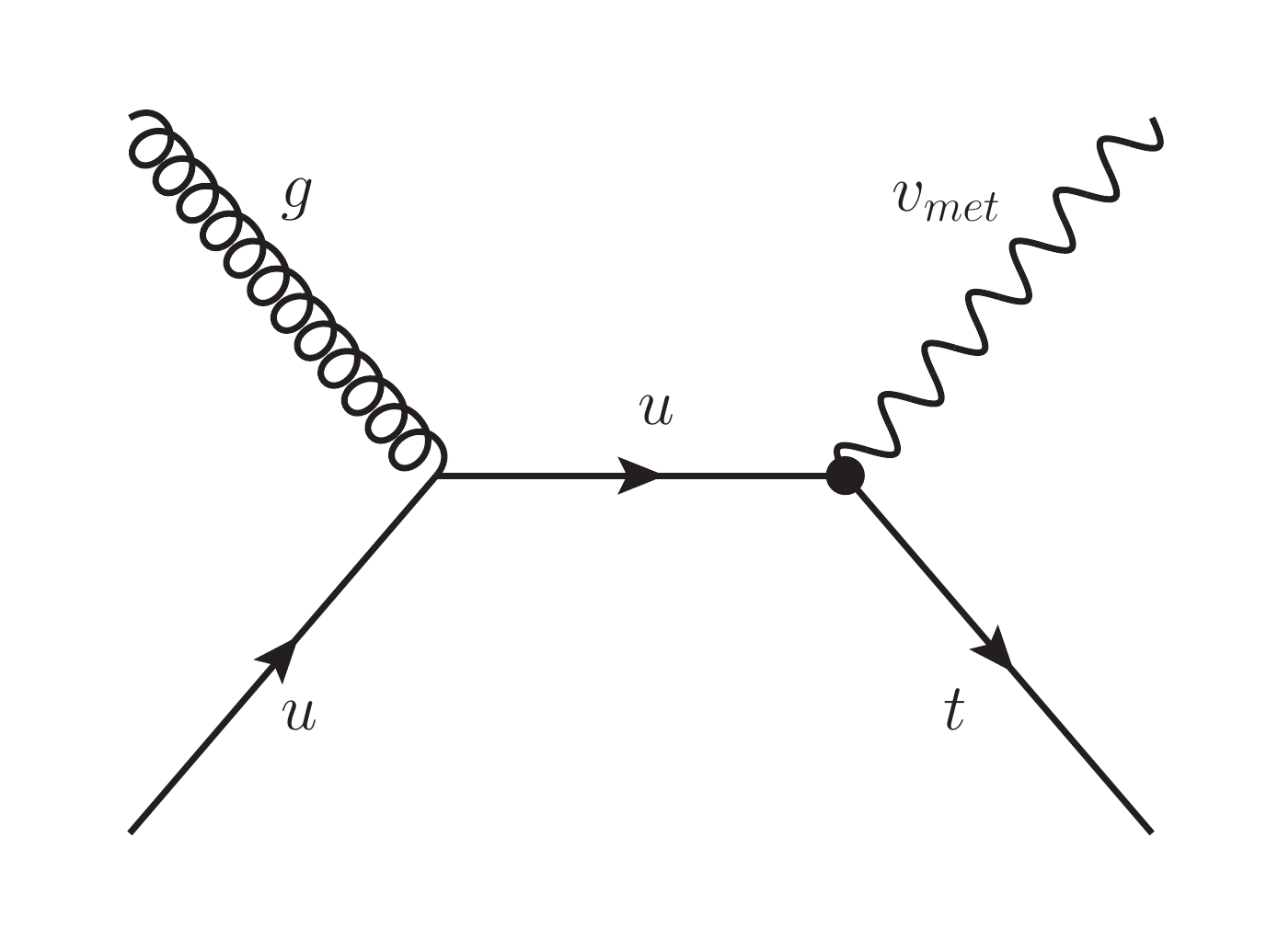}
\includegraphics[width=0.32\textwidth]{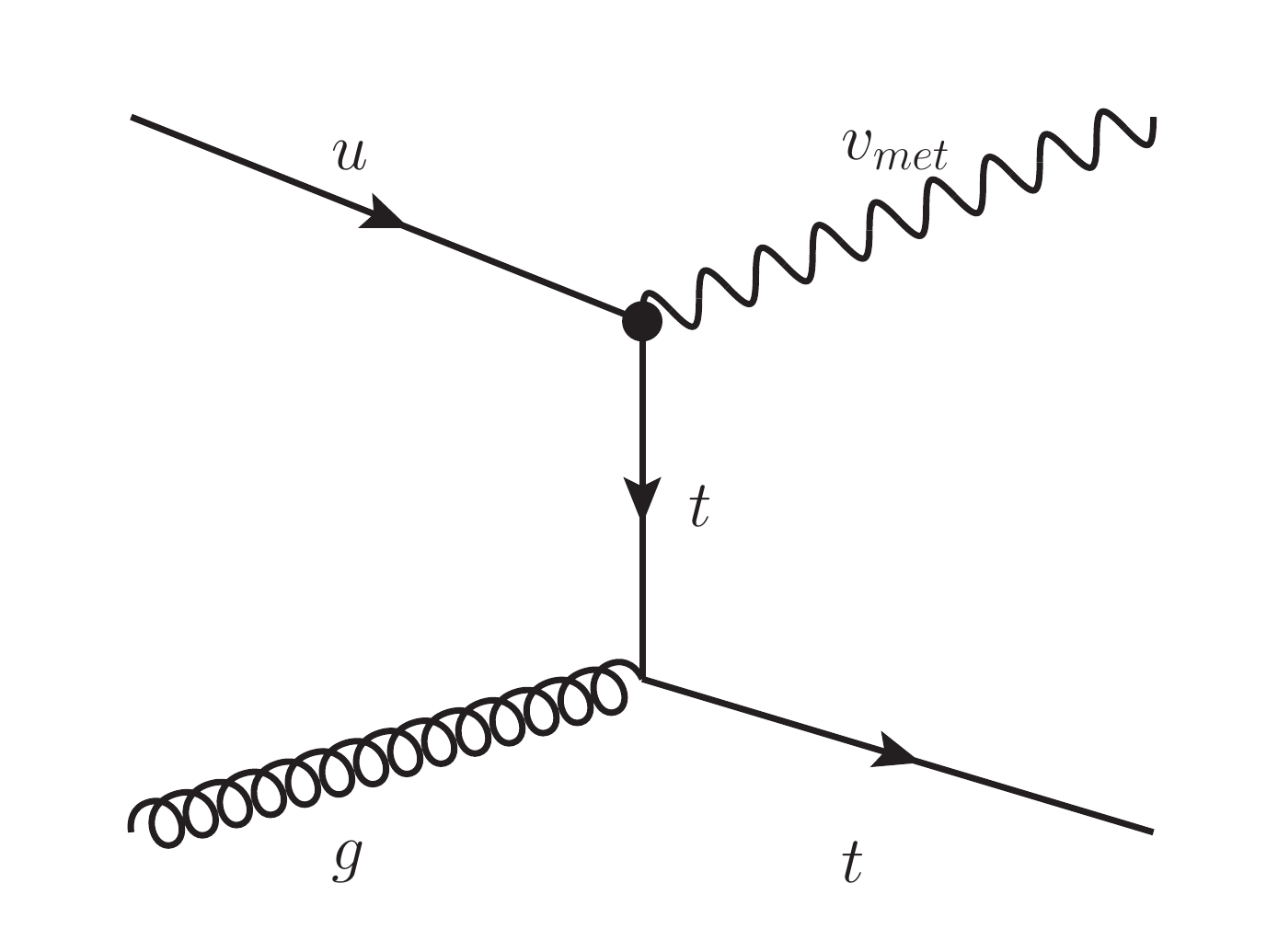}
\caption
{
  Example of Feynman diagrams of leading-order processes leading to monotop events: (left)~production of
  a coloured scalar resonance $S$ decaying into a top quark and a spin-$1/2$ fermion $\fmet$
  in the resonant model, and (middle)~$s$-
  and (right)~$t$-channel non-resonant production of a top quark in association with
  a spin-1 boson $\vmet$ in the non-resonant model.
}
\label{fig:feyn_prod}
\end{figure*}

Because of the variety of these theories, effective models~\cite{AndreaFuksMaltoni,Agrametal} are used for the search reported in this paper.
Furthermore, as minimal extensions of the SM, the effective models tested in this search are required
to respect the electroweak gauge structure~\cite{Cacciapagliaetal}.
The possibilities for monotop production in $pp$ collisions considered are thus:
\begin{itemize}
 \item Resonant production of a +2/3 charged spin-0 boson, $S$, decaying into a right-handed top quark and a neutral, colour singlet, spin-1/2 fermion, $\fmet$;
 \item Non-resonant production of a neutral, colour singlet, spin-1 boson, $\vmet$, in association with a right-handed top quark.
\end{itemize}
The Feynman diagrams for monotop production in the resonant and non-resonant models are shown in Fig.~\ref{fig:feyn_prod}.
Each of these effective models corresponds to one of the two classes of BSM theories detailed above.

A detailed study of the phenomenology of the resonant model is available in Ref.~\cite{Wangetal}.
The interaction Lagrangians of the resonant and non-resonants models are given in Eqs.~(\ref{eq:S1Rlag}) and (\ref{eq:S4Rlag}), respectively.

{\small
\begin{align}
\label{eq:S1Rlag}&{\cal L}_{\mathrm{res}} = \epsilon^{\alpha\beta\gamma} \varphi_{\alpha} \overline{d}^{i,\mathrm{c}}_{\mathrm{\beta,R}} (\ares^q)_{ij} d^j_{\gamma,\mathrm{R}} + \varphi \bar u^{k}_{\mathrm{R}} (\ares^{1/2})_{k} \chi + h.c.\\
\label{eq:S4Rlag}&{\cal L}_{\mathrm{non\mbox{-}res}} = (\anonres)_{ij} V_{\mu} \overline{u}_{\mathrm{R}}^{i} \gamma^{\mu} u_{\mathrm{R}}^{j} +h.c.
\end{align}
}

The fields $\varphi$, $\chi$, and $V_{\mu}$ correspond to the $S$, $\fmet$, and $\vmet$ exotic particles, respectively,
the field $\overline{u}$ ($\overline{d}$) represents an up-type (down-type) quark,
$(\ares^q)_{ij}$, $(\ares^{1/2})_{k}$, and $(\anonres)_{ij}$
are the coupling matrices in the quark-flavour space, the indices $i$, $j$, $k$, represent the quark-generation number,
and $\epsilon^{\alpha\beta\gamma}$ is the fully antisymmetric tensor, the indices $\alpha$, $\beta$, and $\gamma$ being the colour indices.
The superscript $^{\mathrm{c}}$ denotes the charge conjugation.
The number of free parameters is reduced by assuming $(\ares^q)_{\mathrm{12}}=(\ares^q)_{\mathrm{21}}=(\ares^{1/2})_{\mathrm{3}}\equiv \ares$
for the resonant model and $(\anonres)_{\mathrm{13}}=(\anonres)_{\mathrm{31}}\equiv \anonres$ for the non-resonant model,
all other elements of these coupling matrices being equal to 0.
For each model, the coupling parameter $\ares$ or $\anonres$ and the masses of the exotic particles are independent.

The choice of model parameters --- the effective couplings and the masses of the particles --- is driven by phenomenological considerations:
the particles $\fmet$ and $\vmet$ in the resonant and non-resonant models, respectively,
are required to have missing transverse momentum as an experimental signature.
For the resonant model, in which the $\fmet$ fermion can decay into a five-body final state,
Ref.~\cite{Wangetal} suggests that for $m(S)=500$~GeV and an effective coupling of $\ares=0.2$,
the decay length of $\fmet$ is large enough to be considered as invisible for the detector, as long as $\mfmet$ is below 100~GeV.
For the non-resonant model, in which the $\vmet$ boson can decay into a two-body final state either through a tree-level
or a loop-induced interaction, Ref.~\cite{Cacciapagliaetal} assumes that the $\vmet$ boson
decays into a set of invisible particles which can be dark-matter candidates.
This assumption follows the spirit of several BSM models~\cite{Aguilar99,Kamenik,EAlvarezetal}.
Hence, the $\vmet$ particle in the non-resonant model can be considered to be an invisible spin-1 state with mass $\mvmet$.
Studies of possible direct and indirect constraints on monotop model parameters using experimental signatures
other than monotop processes are discussed in Refs.~\cite{Cacciapagliaetal,Wangetal}.

\section{Data and Monte Carlo samples}
\label{sec:samples}

The data used for this analysis are selected from the recorded data streams using
single-electron and single-muon triggers~\cite{ATLAS_triggerperformance}.
Stringent detector and data quality criteria are applied offline, resulting in a data sample corresponding to an
integrated luminosity of \lumierr~\cite{ATLAS_lumi}. 

The signal samples are generated at leading order (LO) in QCD with the matrix-element generator {\sc MadGraph5} v1.5.11~\cite{MG5} using {\sc FeynRules}~\cite{Feynrules1,Feynrules2,Feynrules3}
and interfaced with {\sc Pythia} v8.175~\cite{Pythia6,Pythia8} for parton showering and hadronisation.
The parton distribution function (PDF) set MSTW2008LO~\cite{MSTW2008LO,MSTW2008LO_new} is used.
Resonant signal samples are generated with the mass of the invisible state $\fmet$ varying from 0~GeV to 100~GeV,
the mass of the $S$ resonance being fixed at 500~GeV -- following the suggestion in Ref.~\cite{Wangetal} --
and non-resonant signal samples are generated with the mass of the invisible state $\vmet$ varying from 0~GeV to 1000~GeV.
The couplings $\ares$ and $\anonres$ are set at a fixed value of $0.2$.
In addition, two samples are produced for the resonant model for $\mfmet=100$~GeV,
with coupling strengths fixed at $\ares=0.5$ and $\ares=1.0$,
in order to check the effect of the resonance width on the signal event kinematics. 
The total width of the resonance varies quadratically with the coupling strength,
corresponding to a width of 3.5~GeV, 21.6~GeV, and 86.5~GeV at $\ares=0.2$, $\ares=0.5$, and $\ares=1.0$, respectively.

Top-quark pair  ($t\bar{t}$) and single-top $s$-channel and $Wt$ events are simulated using the next-to-leading order (NLO) generator
{\sc Powheg-Box} v1\_r2129, v1\_r1556, and v1\_r2092, respectively~\cite{powheg1,powheg2}, with CT10 PDF~\cite{CT10,CT10_new}. 
The $t$-channel single-top events are generated using the LO {\sc AcerMC} generator v3.8~\cite{AcerMC_a,AcerMC_b}, with the CTEQ6L1 PDF~\cite{CTEQ6}.
The parton showering, the hadronisation, and the underlying event are modelled using {\sc Pythia} v6.426~\cite{Pythia6}.

The $t\bar{t}$ cross-section for $pp$ collisions at a centre-of-mass energy of
$\sqrt{s} = 8 \tev$ is $\sigma_{t\bar{t}}= 253^{+13}_{-15}$~pb for a top-quark mass of $172.5$~\GeV.
It has been calculated at next-to-next-to-leading order (NNLO) in QCD including resummation
of next-to-next-to-leading logarithmic (NNLL) soft gluon terms~\cite{Cacciari_ttbar,Beneke,Barneuther,Czakon,Mitov,Fiedler} with 
the program Top++ v2.0~\cite{CzakonMitov}.
The PDF and $\alpha_S$ uncertainties were calculated using the PDF4LHC prescription~\cite{PDF4LHC}
with the MSTW2008 68\% CL NNLO~\cite{MSTW2008LO,MSTW2008LO_new}, CT10 NNLO~\cite{CT10,CT10_new}
and NNPDF2.3 5f FFN~\cite{NNPDF23_new} PDF sets, added in quadrature to the scale uncertainty.
The single-top cross-sections are obtained from approximate NNLO calculations:
87.8$^{+3.4}_{-1.9}$~pb ($t$-channel), 22.4$\pm$1.5~pb ($Wt$ process)
and 5.6$\pm$0.2~pb ($s$-channel)~\cite{Kidonakis_tchan,Kidonakis_Wt,Kidonakis_schan}.

The {\sc Alpgen} LO generator v2.14~\cite{Alpgen} is used with {\sc Pythia} v6.426
to generate events with a $W$ boson produced in association with light or heavy quarks
($W$+light-quarks, $W$+$bb$, $W$+$cc$, $W$+$c$) and $Z$+jets events.
The {\sc Alpgen} matrix elements include diagrams with up to five additional partons.
To remove overlaps between the $n$ and $n+1$ parton samples
the MLM matching scheme~\cite{MLM} is used.
Double counting between the $W$+$n$ parton samples and
samples with associated heavy-quark pair production is removed utilising
an overlap removal based on a $\Delta R=\sqrt{(\Delta\eta)^2+(\Delta\phi)^2}$ matching criterion.
Diboson samples ($WW$, $ZZ$, and $WZ$) where at least one of the bosons decays leptonically are modelled by {\sc Herwig} v6.52~\cite{SAMPLES-HERWIG}. 
The single-boson and diboson simulation samples are normalised to the production cross-sections calculated
at NNLO~\cite{Hamberg,FEWZ} and NLO~\cite{MCFMpaper}, respectively.

After event generation, all signal and background samples are passed through the full simulation
of the ATLAS detector~\cite{ATLASsim} based on GEANT4~\cite{SAMPLES-G4}
and reconstructed using the same procedure as for collision data.
All Monte Carlo (MC) samples are simulated with pile-up\footnote{Extra proton--proton interactions from the same and previous bunch-crossings.} 
and re-weighted to have the same distribution of the mean number of interactions per bunch-crossing as in the data sample (20.7 on average).

%
\section{Selection and background estimation}
\label{sec:objects}

The experimental signature of the monotop events is one isolated charged lepton (electron or muon) from the $W$ decay,
large missing transverse momentum, and one jet identified as likely to have originated from a $b$-quark ($b$-tagged).

Electrons are identified as energy clusters in the electromagnetic calorimeter matched to reconstructed tracks 
in the inner detector~\cite{ele_calib,Electron2010,Egamma2011Paper}.
Electron candidates are required to be isolated from other objects in the event and from hadronic activity to reduce the contamination 
by mis-reconstructed hadrons, electrons from heavy-flavour decays, and photon conversions.
Muons are reconstructed using information from the muon spectrometer and the inner detector~\cite{MuonPerf}.  
An isolation criterion~\cite{MiniIso} is applied to reduce the contribution of muons from heavy-flavour decays.
The reconstructed charged lepton is required to have a transverse momentum \pt~$>$ 30~\GeV~to ensure a constant 
trigger efficiency and to have $|\eta|<$ 2.5 for muons and $|\eta|<$ 2.47 for electrons 
(for the latter, the electromagmetic calorimeter barrel--endcap transition region 1.37 $<|\eta|<$ 1.52 is excluded).

Jets are reconstructed using the anti-\kt\ algorithm~\cite{KtAlgo} with a radius parameter $R=0.4$
and calibrated to the hadronic energy scale~\cite{JES_paper}. 
Jets are required to have \pt\ $> 25$~\gev\ and $\left| \eta \right| < 2.5$.
To suppress jets from in-time pileup, at least 50\% of the scalar \pT~sum of the tracks associated with a jet 
is required to be from tracks associated with the primary vertex. 
This jet vertex fraction requirement is applied only for jets with \pt~$<50$~\gev\ and $|\eta|<2.4$.

Exactly one jet is selected, and is required to be $b$-tagged.
The $b$-tagging techniques are based on properties specific to $b$-hadrons, such as long lifetime and large mass~\cite{btag}.
This analysis uses a neural-network-based $b$-tagger which combines several $b$-tagging algorithms. 
The chosen working point corresponds to a $b$-tagging efficiency of 57\% and a light-quark selection efficiency of 0.2\%, 
as obtained in simulated $t\bar{t}$ events.

The missing transverse momentum (with magnitude \MET) is the negative vector sum of the transverse momentum
associated with topological clusters of energy deposits in calorimeter cells
and is further refined with object-level corrections from identified electrons, muons, and jets~\cite{ETmissresol1,ETmissresol2}.
This analysis requires events to have \met\ larger than 35~\gev\ to reduce the multijet background.
\begin{figure*}[!htb]
\centering
\includegraphics[width=0.45\textwidth]{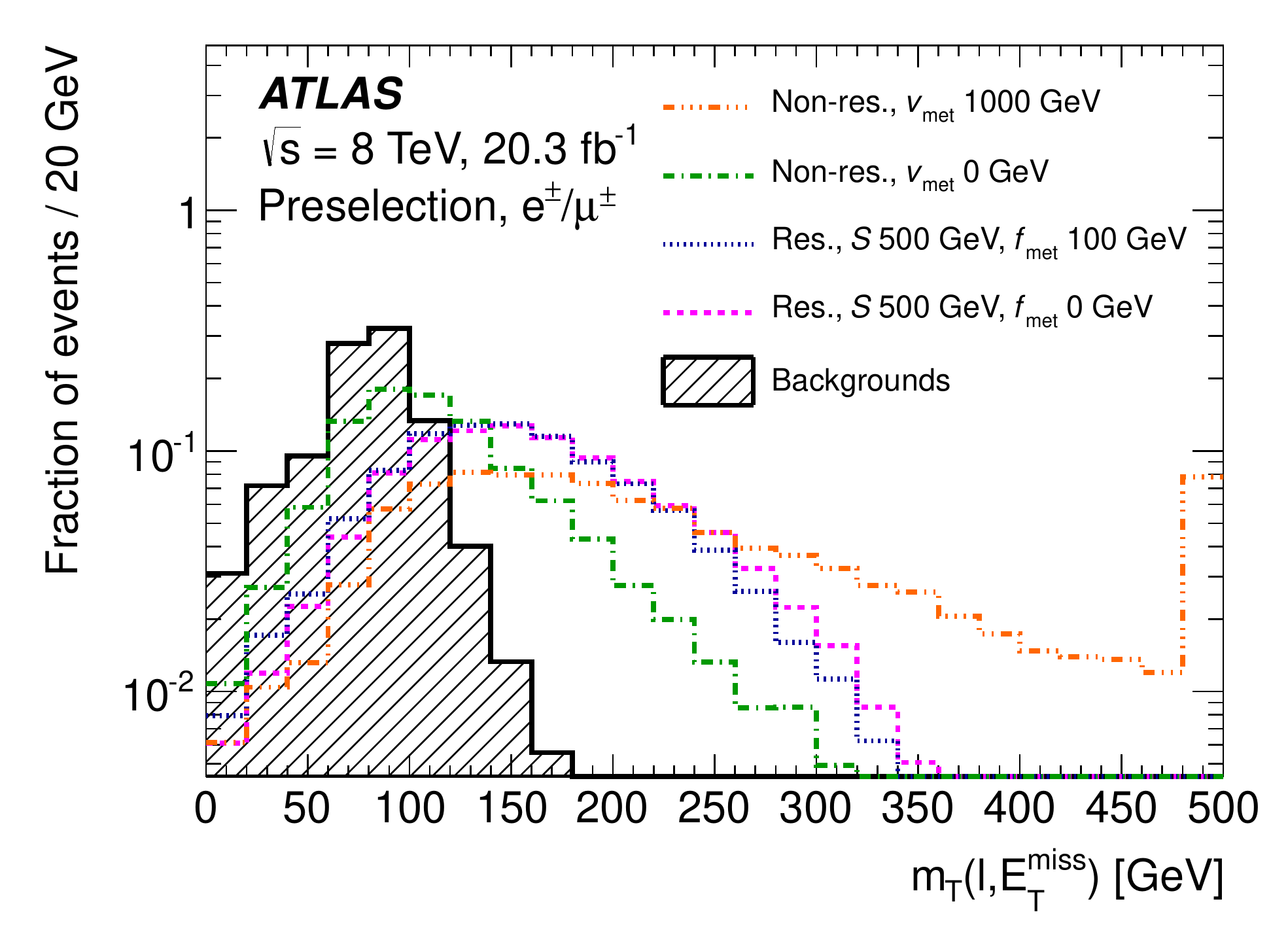}
\includegraphics[width=0.45\textwidth]{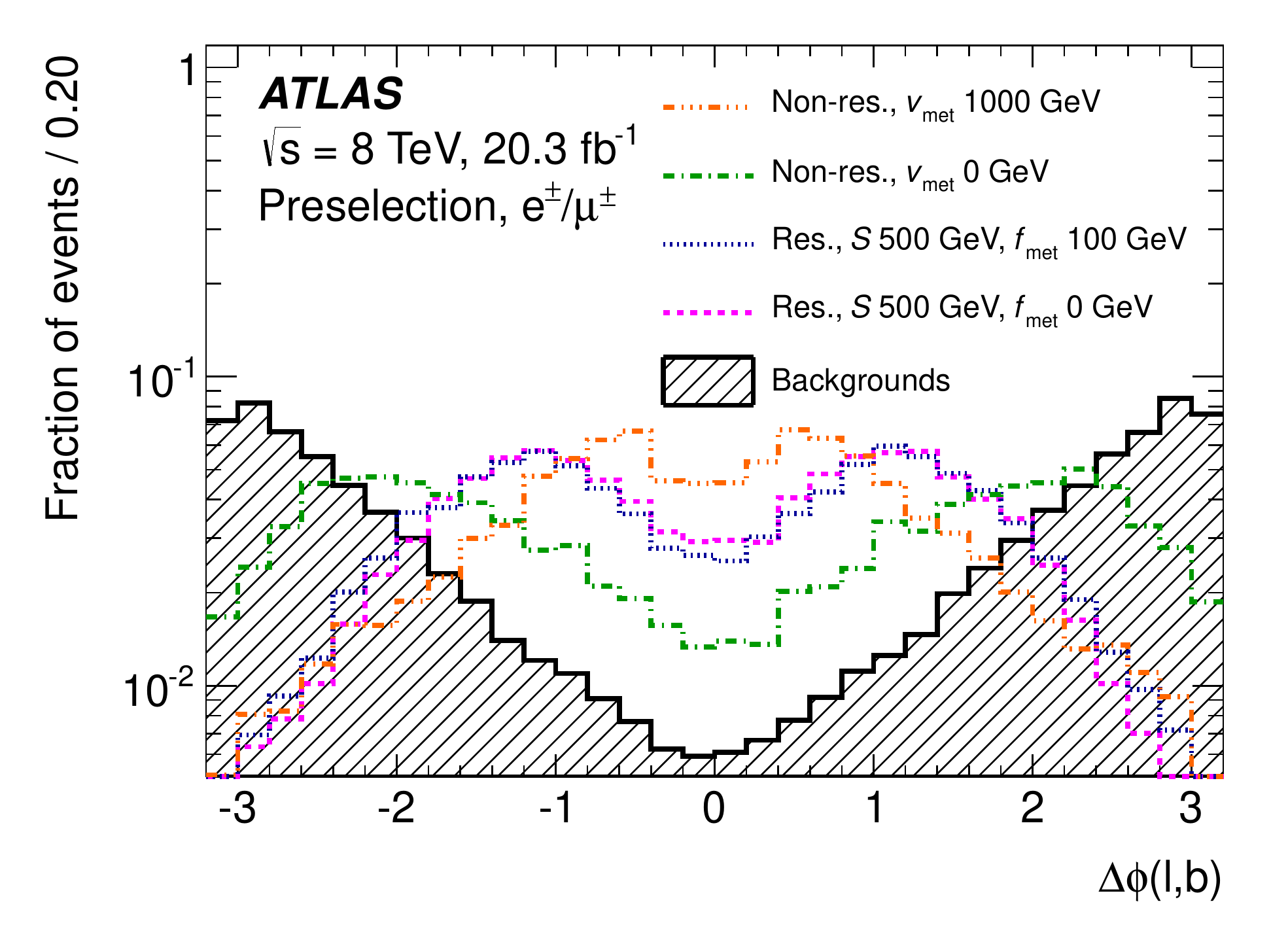}
\caption
{
  Distributions normalised to unity of (left) $\MTW$ and of (right) $\deltaphilb$
  for events satisfying the pre-selection defined in the text.
  The expected distributions for the resonant model with $m(S)=500$~GeV are shown for the $\mfmet=0$~GeV and $\mfmet=100$~GeV
  hypotheses, as well as for the non-resonant model for the $\mvmet=0$~GeV and $\mvmet=1000$~GeV hypotheses.
  All distributions are compared to the expected distribution for the backgrounds.
  For the $\MTW$ distributions, the last bin includes overflows.
}
\label{fig:BkgSignalComparison}
\end{figure*}

The main background to this final-state selection are \ttbar\ pairs where both top quarks decay semi-leptonically, $t\to \ell\nu b$,
with large \MET\ due to one lepton and one jet not being reconstructed,
and $W$+jets production, particularly with jets from heavy-flavour quarks.
The background from multijet production due to misidentification as leptons is
reduced by imposing a requirement on the sum of the \MET~and the transverse mass\footnote{The transverse mass is defined as\\$\MTW=\sqrt{2 \pT\left(\ell\right) \MET \left(1-\cos \Delta \phi \left(\pT\left(\ell\right), \MET\right) \right)}$,
where $\pT\left(\ell\right)$ denotes the modulus of the lepton transverse momentum, and $\Delta \phi \left(\pT\left(\ell\right), \MET\right)$ the azimuthal difference between the missing transverse momentum and the lepton directions.}  of the lepton--\MET\ system:
$\MTW + \MET>~60~\mathrm{\GeV}$.
The distributions of kinematic variables and their normalisation for the multijet background are estimated
with a data-driven matrix method~\cite{MatrixMethod}. 
All remaining background processes (\ttbar, single-top, $W$+jets, $Z$+jets and diboson production) are modelled 
using simulated samples and are scaled to the theory predictions described in Sect.~\ref{sec:samples}.
Possible contributions from $t\bar{t}Z$ and $tZ$ processes~\cite{tZCrossSections} in the $Z\rightarrow\nu\nu$ decay mode
are found to be negligible.

A counting experiment approach is followed.
The monotop signal is prominent in regions of the phase space characterised by high $\MTW$ values, as suggested by Refs.~\cite{EAlvarezetal,Agrametal}.
Hence, in addition to the pre-selection described previously,
a criterion requiring $\MTW>150~\mathrm{\GeV}$ is used to define the signal region.
In order to improve the sensitivity of the search, an optimisation of the event selection
is performed with simulated data, using well-modelled variables.
The lepton and the $b$-tagged jet are closer to each other when originating
from the decay of a top quark than in the case of $W$+jets and multijet background events.
Hence, a criterion imposing the rejection of events with large values of the difference in azimuth between the lepton and the $b$-tagged jet
$|\deltaphilb|$ is tested, together with increased $\MTW$ threshold values.
Figure~\ref{fig:BkgSignalComparison} shows the distributions of these two variables for the expected background contribution, and for two mass hypotheses
considered for each signal model.
For each set of cuts on $\MTW$ and $|\deltaphilb|$, the sensitivity is estimated
by calculating the expected limit on the production cross-section with the procedure described in Sect.~\ref{sec:results}
including the systematic uncertainties detailed in Sect.~\ref{sec:systematics}.
The optimisation was performed using one mass hypothesis $\mfmet=\mathrm{100~GeV}$ for the resonant model,
for which the kinematic distributions have only small variations in the studied mass range.
For the non-resonant model, characterised by larger variations of the kinematic distributions with $\vmet$,
four signal mass hypotheses were studied: $\mvmet=0$~GeV, 100~GeV, 300~GeV, and 600~GeV.
The resulting best-performing selections, for the tested mass hypotheses, are:
\begin{itemize}
  \item SRI (resonant model optimisation):\\$\MTW>\mathrm{210~GeV}$ and $|\deltaphilb|<1.2$
  \item SRII (non-resonant model optimisation):\\$\MTW>\mathrm{250~GeV}$ and $|\deltaphilb|<1.4$
\end{itemize}

\begin{figure}[!htb]
\centering
\includegraphics[width=0.4\textwidth]{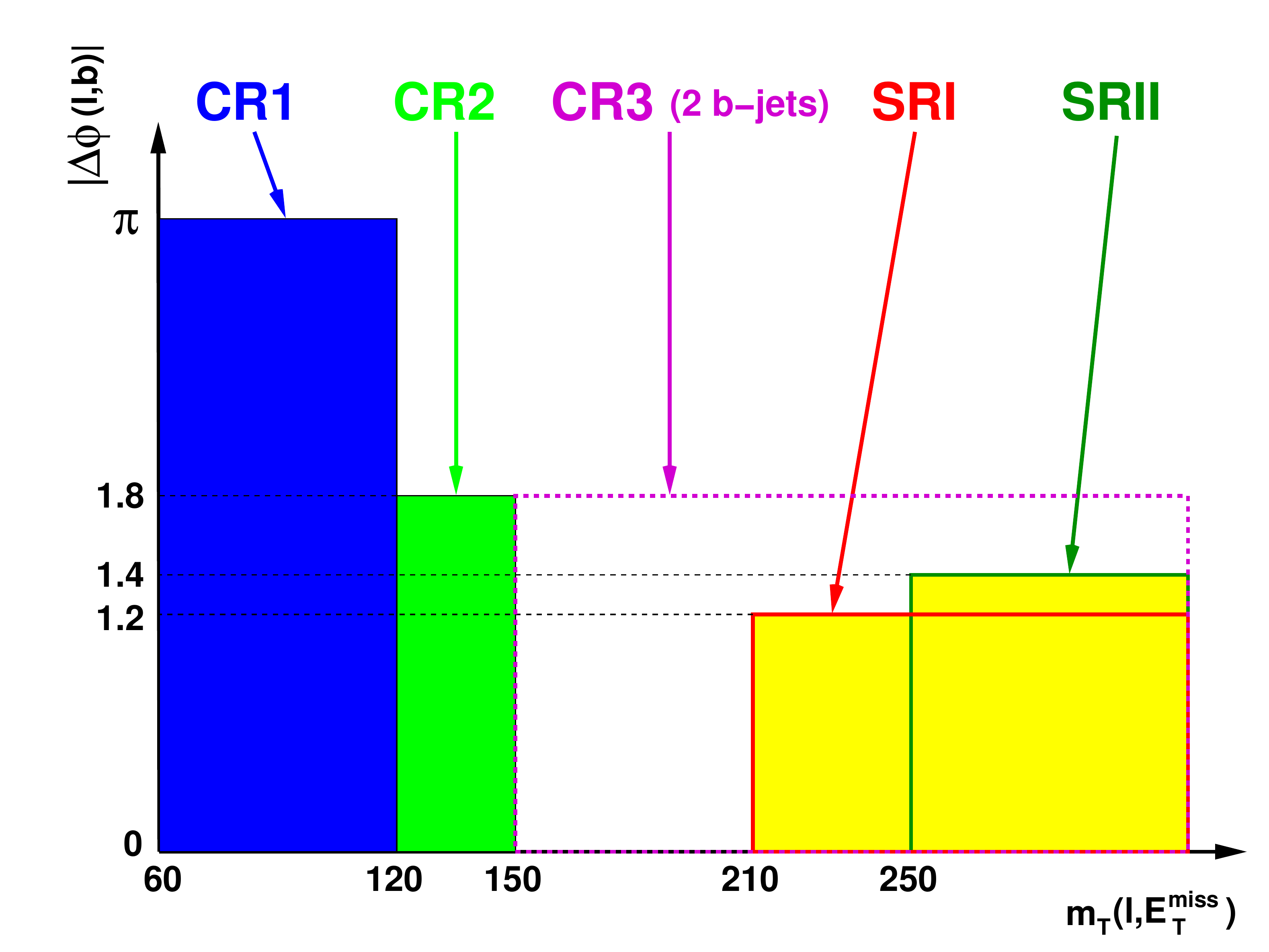}
\caption
{
  Sketch depicting the control and signal regions in the ($\MTW$, $|\deltaphilb|$)-space.
  The CR1 (CR2) control region is defined as $60\mathrm{~GeV}<\MTW<120\mathrm{~GeV}$
  ($120\mathrm{~GeV}<\MTW<150\mathrm{~GeV}$ and $|\deltaphilb|<1.8$).
  The CR3 control region corresponds to $\MTW>150\mathrm{~GeV}$ and $|\deltaphilb|<1.8$, but with a second $b$-tagged jet.
  The SRI (SRII) signal selection optimised for the resonant (non-resonant) signal model, is defined as $\MTW>\mathrm{210~GeV}$ and $|\deltaphilb|<1.2$
  ($\MTW>\mathrm{250~GeV}$ and $|\deltaphilb|<1.4$).
}
\label{fig:CRSkecth}
\end{figure}

\begin{figure*}[p]
\centering
\includegraphics[width=0.4\textwidth]{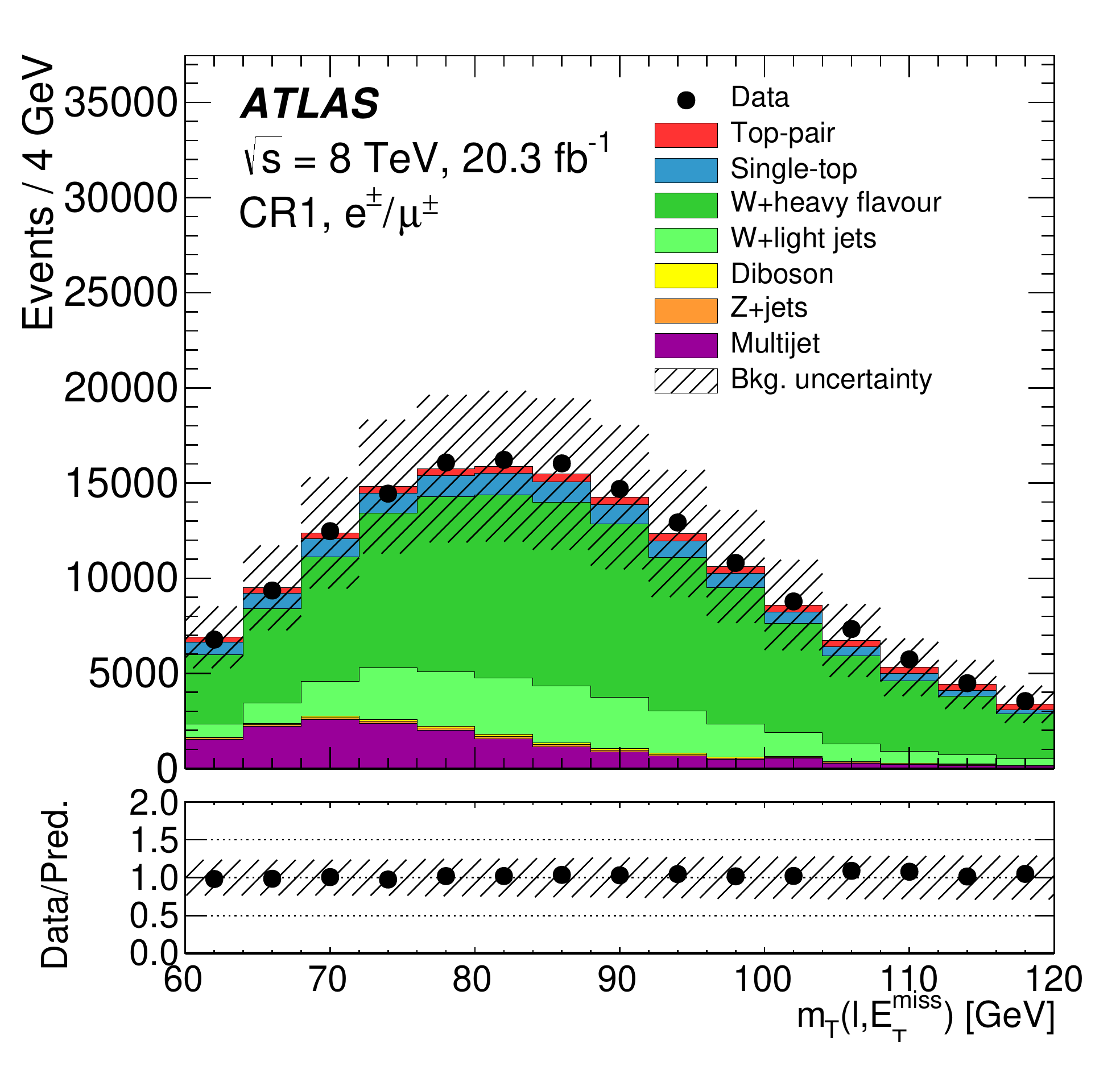}
\includegraphics[width=0.4\textwidth]{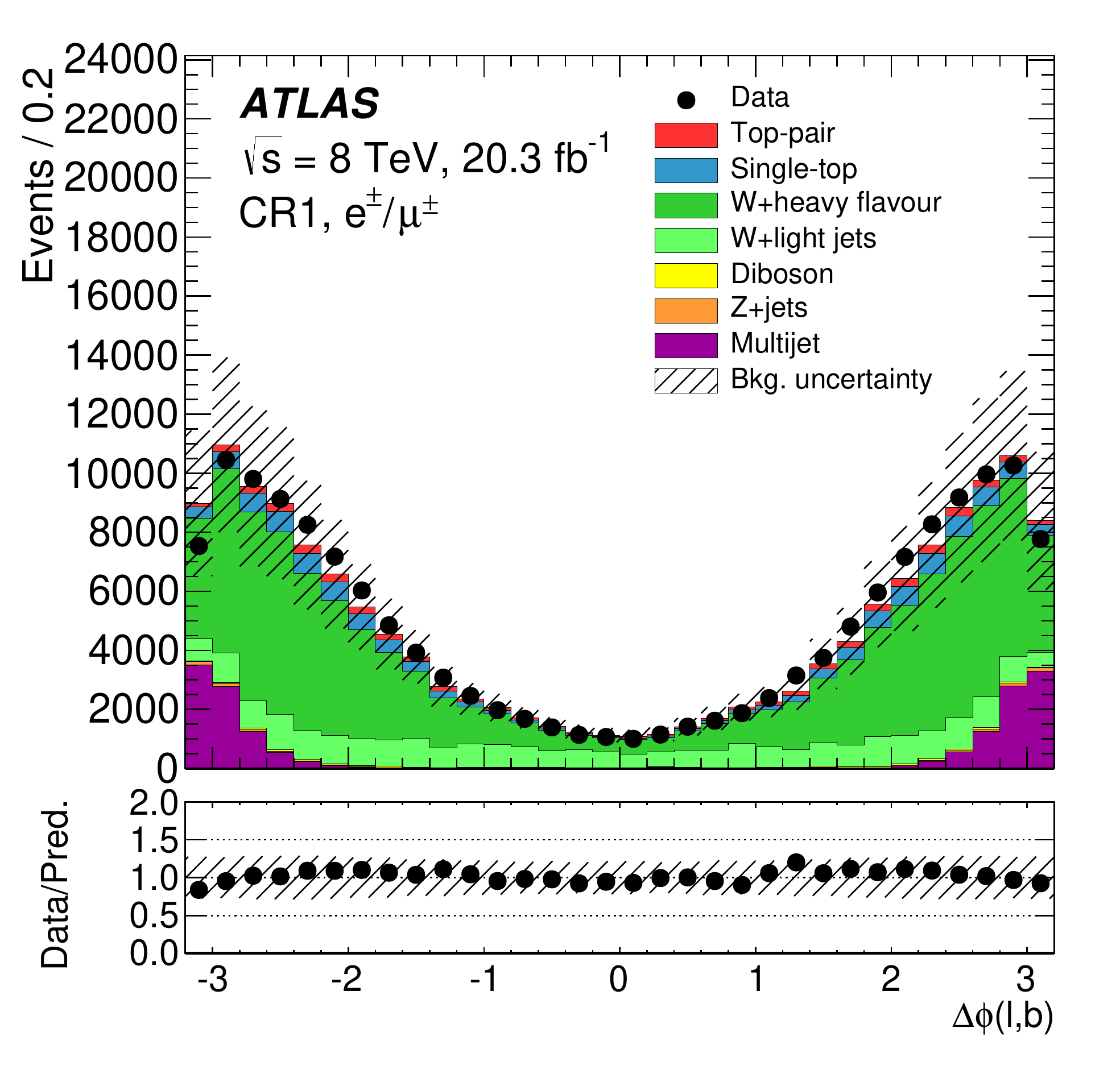}\\
\includegraphics[width=0.4\textwidth]{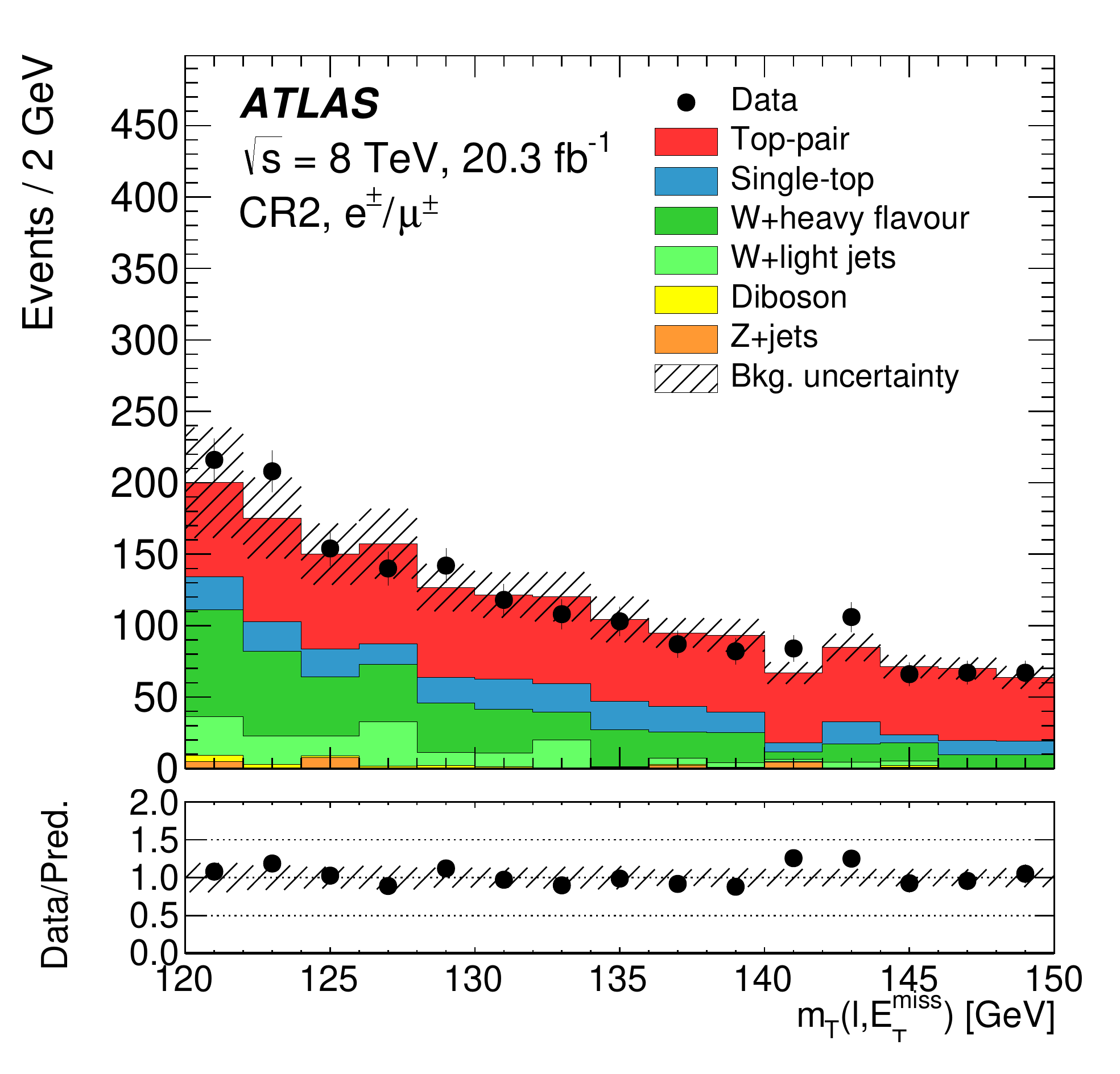}
\includegraphics[width=0.4\textwidth]{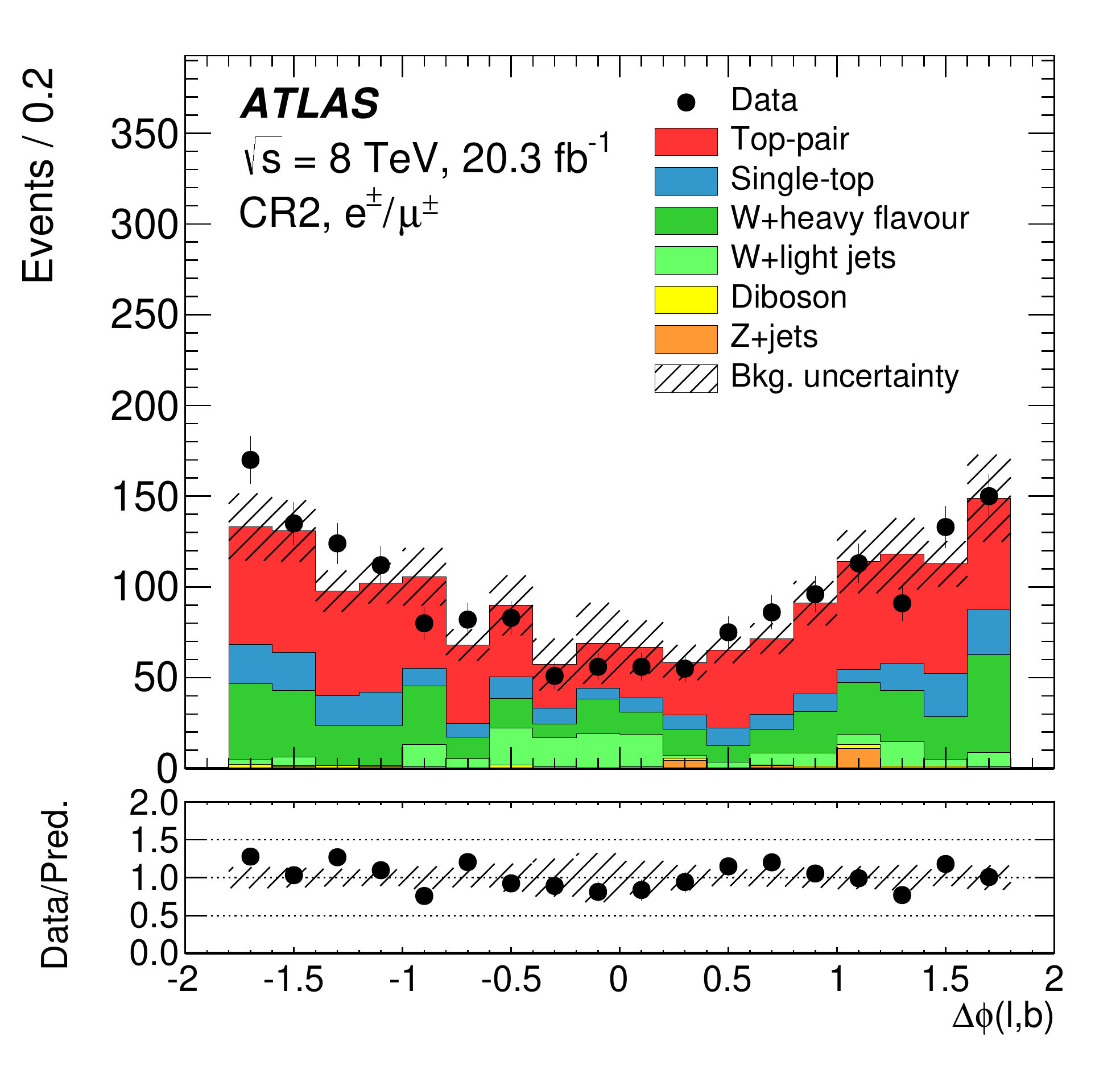}\\
\includegraphics[width=0.4\textwidth]{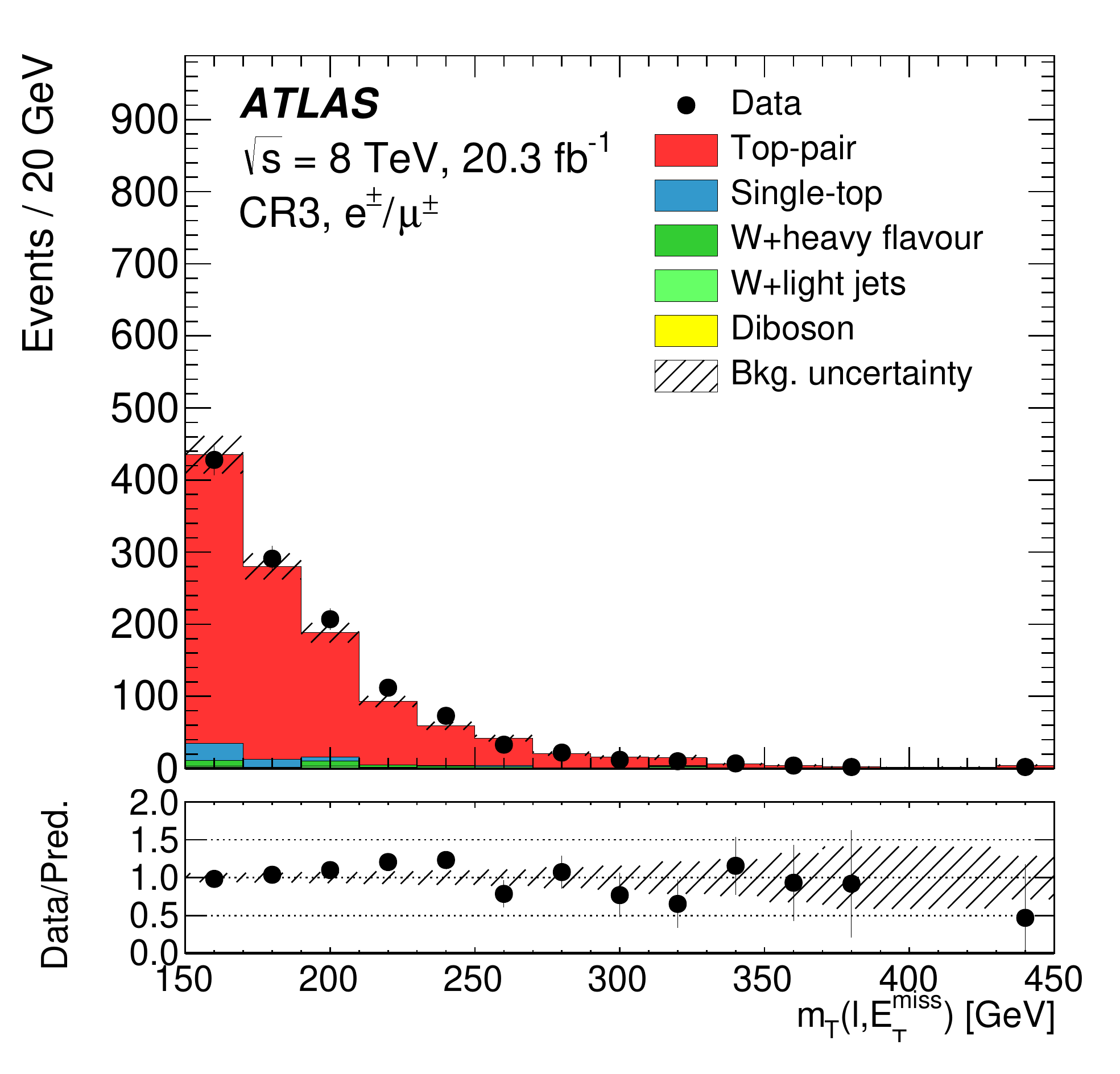}
\includegraphics[width=0.4\textwidth]{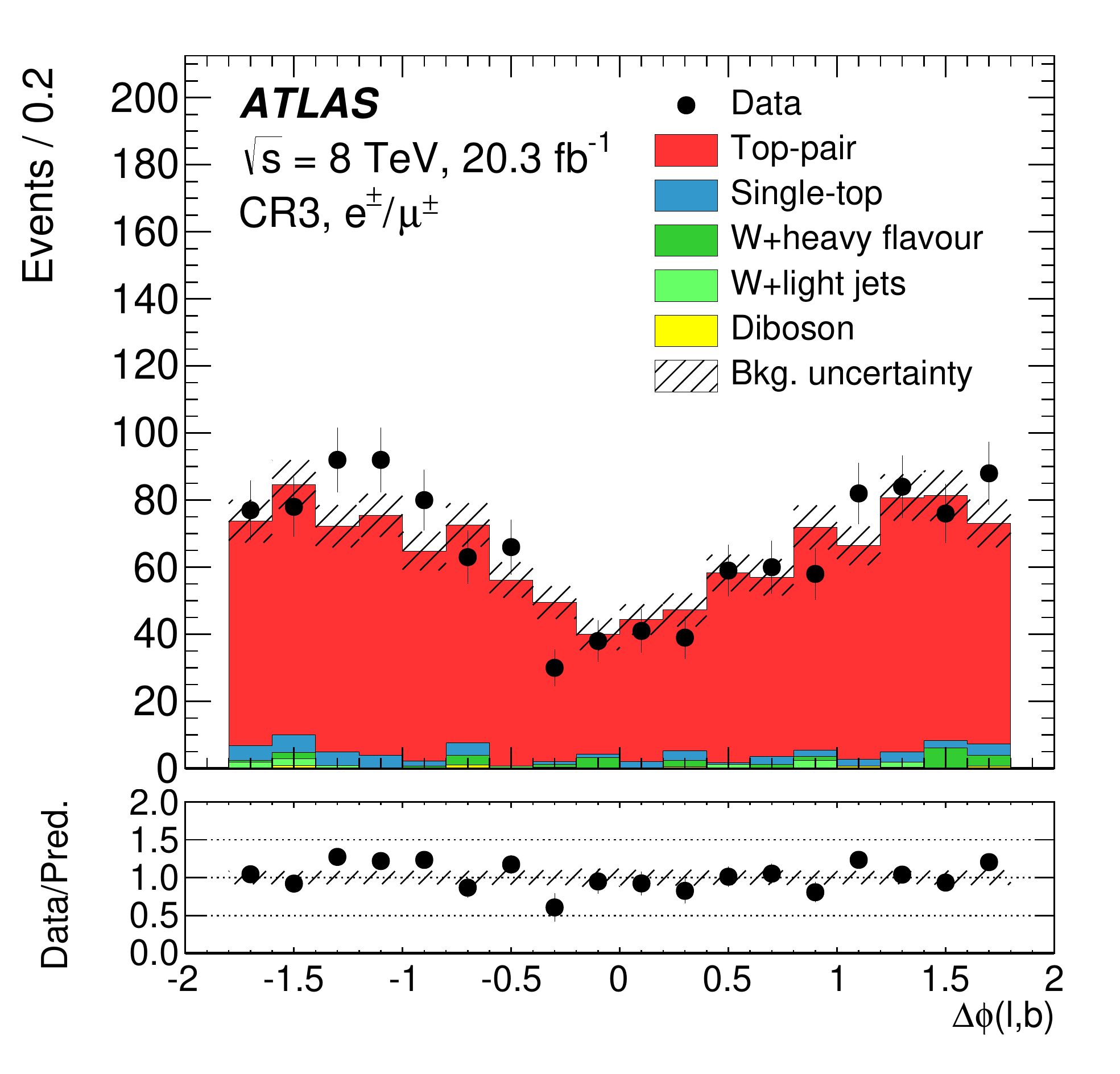}
\caption
{
  Distributions of (left) $\MTW$ and of (right) $\deltaphilb$ in (top) the CR1, (middle) the CR2,
  and (bottom) the CR3 control region, for the electron and muon channels combined.
  The distributions observed in data, depicted with the points, are compared with the predicted background contributions.
  In the CR2 and CR3 regions, the negligible multijet contribution is not shown, and neither is the $Z$+jets contribution in the CR3 region.
  The multijet background is normalised by the data-driven method, and the other backgrounds are normalised to their theoretical cross-sections.
  The error bands correspond to the uncertainties due to the statistical uncertainty of the sample added in quadrature with a conservative 50\% normalisation uncertainty on
  the multijet contribution, and with the $W$+jets and \ttbar\ cross-section uncertainties.
  The ratios of the observed distributions to the predicted background distributions are shown in the lower frame.
}
\label{fig:CRplots}
\end{figure*}

In order to validate the background model, three control regions orthogonal to the signal region are defined.
Figure~\ref{fig:CRSkecth} is a sketch describing the signal and control regions in the ($\MTW$, $|\deltaphilb|$)-plane.
The first control region (CR1) is enriched in $W$+jets and multijet background events by requiring events to satisfy $60\mathrm{~GeV}<\MTW<120\mathrm{~GeV}$
in addition to the pre-selection criteria.
In the second control region (CR2) with a kinematic regime closer to the one of the signal region,
the pre-selected events are required to satisfy $120\mathrm{~GeV}<\MTW<150\mathrm{~GeV}$ 
and the azimuthal separation $\deltaphilb$ between the lepton and the $b$-tagged jet must be less than $1.8$.
Finally, the third control region (CR3) is defined in order to validate the modelling of the background arising from \ttbar~events.
An event sample dominated by \ttbar~events is obtained by selecting events with a second $b$-tagged jet;
both $b$-tagged jets are identified with a $b$-tagging criterion with an efficiency of 80\%, the sub-leading jet satisfies $\pt<50$~GeV,
and the events must satisfy $\MTW>150\mathrm{~GeV}$ and $|\deltaphilb|<$~1.8 in addition to the pre-selection criteria.
The distributions of $\MTW$ and of $\deltaphilb$ in the three control regions are depicted in Fig.~\ref{fig:CRplots}.
Reasonable agreement between the data and the predicted background estimate is found.

\section{Systematic uncertainties}
\label{sec:systematics}

The impact of systematic uncertainties is considered
on the yields of individual background and signal processes.
The main systematic uncertainties are those related to the jet energy scale,
the $b$-tagging efficiency, the effect of the choice of PDF on signal and background acceptance,
the effect of the choice of MC generator and of additional radiation on \ttbar\ modelling,
and the effect of the limited size of the samples.

\subsection{Sample size}
\label{subsec:systMCStat}

Due to the stringent kinematic cuts in the signal regions,
the impact of the limited size of the data and simulated samples on the signal and background estimates is a significant source of systematic uncertainty.
For the $Z$+jets, multijet, and single-top-quark $s$- and $t$-channel processes, the expected event yield is zero in both channels,
for the SRI and SRII selections, respectively.
In such cases, a 68\%~confidence level (CL) upper limit on the yields is calculated, assuming a Poisson distribution, and is taken into account in the limit-setting procedure.
This upper limit represents at most 10\% of the background contribution.

For the other processes, which have non-negligible contributions,
the effect of the limited sample size on expected signal (background) yields varies between 2\% and 5\% (2 and 9\%).

\subsection{Object modelling}
\label{subsec:systObjects}

The effect of the  uncertainty on the jet energy scale~\cite{JES_paper} is a change in the signal (background) event yields of 1--5\% (9--10\%),
depending on the channel and on the signal region.
The impact of the jet energy resolution uncertainty, evaluated by smearing the jet energy in the simulation~\cite{JER},
is a 2--3\% (1--2\%) effect on the signal (background) rates.
The systematic uncertainty associated with the efficiency of the cut on the jet vertex fraction
results in yield variations of 2--3\% (2--6\%) in the signal (background).
Uncertainties on $b$-tagging efficiency and mistagging rates are estimated from data~\cite{btag};
the effect on signal and background yields is 3--5\%.
The jet reconstruction efficiency uncertainty has an effect below 1\%, except for the background in the SRII region (up to 3\%).

Smaller uncertainties arise from the lepton trigger, reconstruction, and identification efficiencies (up to 1\%)
and from lepton energy scale and resolution (up to 1\% for signal and between 1\% and 3\% for background).
The systematic uncertainties related to leptons and jets are propagated to the \MET.
In addition, uncertainties on the estimation of the contributions of calorimeter energy deposits
not associated with any reconstructed objects have an effect below 1\% (up to 4\%) on expected signal (background) contribution.

\subsection{Signal and background acceptance modelling}
\label{subsec:systAccModelling}

The uncertainties on the signal and background acceptance due to the choice of PDF are estimated using the CT10~\cite{CT10,CT10_new},
MSTW2008 68\% CL NLO~\cite{MSTW2008LO,MSTW2008LO_new} and NNPDF2.3~\cite{NNPDF23,NNPDF23_new} PDF sets with their uncertainties,
following the PDF4LHC recommendations~\cite{PDF4LHC}.
The variations of the signal (background) yields are between 4--11\% (5--6\%).

The dependence of the \ttbar\ process on the generator and parton showering
simulation is evaluated by comparing the nominal sample produced with {\sc Powheg+Pythia} with three samples generated using the CT10 PDF,
one sample produced with {\sc Powheg-Box} v1\_r2129, one sample using the {\sc Alpgen} LO multileg generator v2.14~\cite{Alpgen},
and one sample produced using {\sc MC@NLO} v4.06~\cite{McAtNlo_a,McAtNlo_b}.
{\sc Herwig} v6.52~\cite{SAMPLES-HERWIG} is used for parton showering and hadronisation and {\sc Jimmy} v4.31~\cite{Jimmy} for the underlying event.
The largest variation, representing 5--11\% of the total background yield, arises from the comparison with the {\sc Alpgen+Herwig} sample.
For $Wt$ production, the nominal {\sc Powheg+Pythia} sample is compared with a sample produced with {\sc MC@NLO} v4.06,
leading to a variation of 4--6\% on the total background yield.
Furthermore, the uncertainty associated with the NLO calculation schemes for the $Wt$ process
is evaluated by comparing the nominal sample generated with the diagram removal scheme
to a sample using the diagram subtraction (DS) scheme~\cite{WtProd};
this uncertainty is 3--5\% on the total background yield.

The dependence of the \ttbar\ event rate on additional radiation is evaluated using a \ttbar\ sample generated with the
{\sc AcerMC} LO generator v3.8~\cite{AcerMC_a,AcerMC_b}, with the CTEQ6L1 PDF set~\cite{CTEQ6}, and
coupled with {\sc Pythia} v6.426.  The {\sc Pythia} parameters are varied in a manner consistent with
a measurement of $t\bar{t}$ production with additional jet activity~\cite{ttbarveto}.
The related variation in the total background is around 5\% (9\%) in the SRI (SRII) region.

\subsection{Background normalisation}
\label{subsec:systBkgNorm}

Theoretical uncertainties are $-$5.9/+5.1\% for the inclusive $t\bar{t}$ cross-section~\cite{Cacciari_ttbar,Beneke,Barneuther,Czakon,Mitov,Fiedler,CzakonMitov}, 
and 6.8\% for the $Wt$-channel cross-section~\cite{Kidonakis_Wt}.
An uncertainty of 24.5\% on diboson and $W$+light-quarks rates is also assigned.
These estimates come from the uncertainty on the inclusive diboson and $W$-boson production cross-sections~\cite{FEWZ} (5\% and 4\% respectively)
and from a conservative assessment based on a prediction for the ratio of the event rate with $n+1$ jets
to the event rate with $n$ jets~\cite{Berger,Berends1}, resulting in 24\% per additional jet, added in quadrature.
A 50\% uncertainty, as evaluated in Ref.~\cite{tchanXsec7TeV}, is assigned to the $W$+$bb$, $W$+$cc$, and $W$+$c$ rates.

\subsection{Luminosity}
\label{subsec:systLumi}

The uncertainty on the integrated luminosity is 2.8\%~\cite{ATLAS_lumi},
affecting the signal estimates as well as the simulated backgrounds.

%
\section{Results and interpretation}
\label{sec:results}

\begin{figure*}[!htb]
\centering
\includegraphics[width=0.44\textwidth]{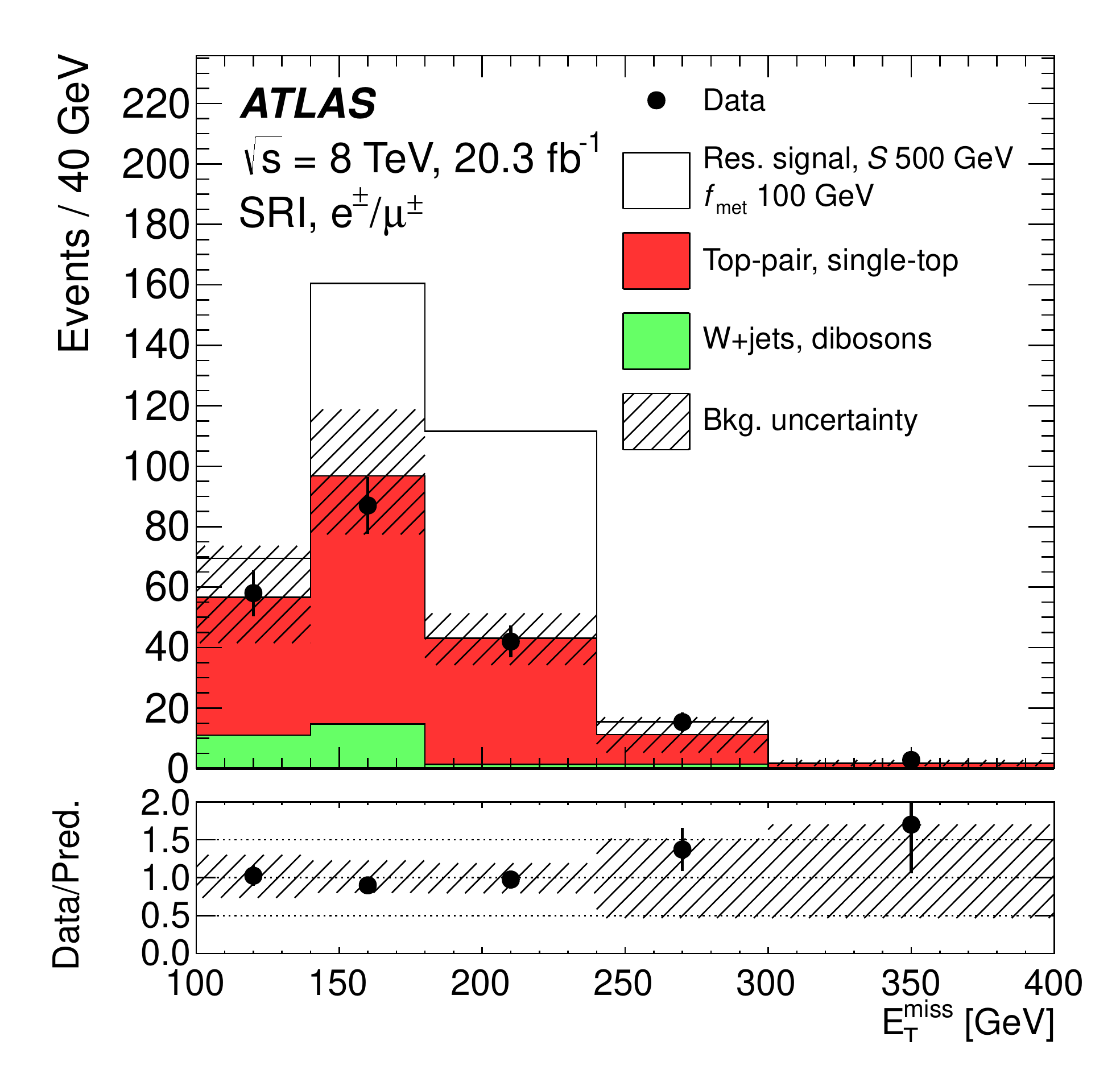}
\includegraphics[width=0.44\textwidth]{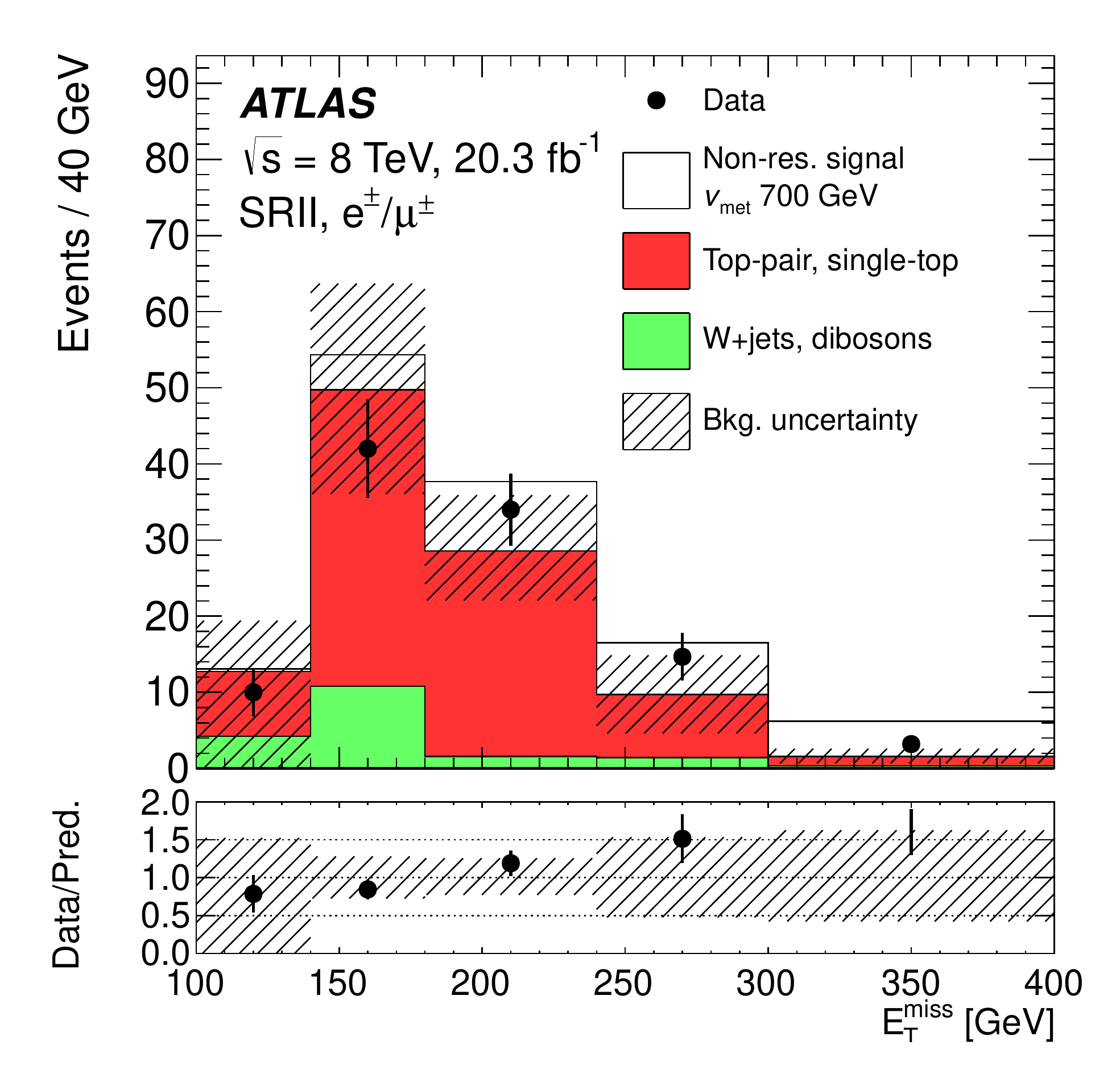}
\caption
{
  Distributions of \MET\ in the (left) SRI and (right) SRII signal regions,
  for the electron and muon channels combined.
  The distributions observed in data, depicted with the points, are compared with the predicted background contributions,
  shown stacked together with the expected resonant (non-resonant) signal contribution for the $\mfmet=100$~GeV and $m(S)=500$~GeV
  ($\mvmet=700$~GeV) hypothesis.
  The expected backgrounds are normalised to their theoretical cross-sections, and the expected resonant (non-resonant) signal
  is normalised to the theoretical cross-section corresponding to $\ares=0.2$ ($\anonres=0.2$).
  The error bands on the expected backgrounds correspond to the uncertainties due to all systematic sources added in quadrature.
  The first (last) bin includes underflows (overflows).
  The ratios of the observed distributions to the predicted background distributions are shown in the lower frame.
}
\label{fig:DataMCPlots}
\end{figure*}

Figure~\ref{fig:DataMCPlots} shows the distributions of \MET\ in the SRI and SRII signal regions,
comparing the data to the expected signal and background contributions.
The expected resonant (non-resonant) signal contribution for the $\mfmet=100$~GeV
($\mvmet=700$~GeV) hypothesis, normalised to the theoretical cross-section corresponding
to $\ares=0.2$ ($\anonres=0.2$), is also shown.

Table~\ref{tab:Yields} reports the expected event yields for the background and signal processes
and the observed event yields in the SRI and SRII signal regions.
As no excess is observed in data, 95\%~CL upper limits on the signal production cross-sections
are set with the $CL_s$ procedure~\cite{CLsJunk,CLsAlexander}.
A log-likelihood ratio (LLR) is used as the test statistic, defined as the ratio of the
signal-plus-background hypothesis to the background-only hypothesis.
For a given hypothesis, the combined likelihood is the product of the likelihoods for the two channels considered (electron and muon),
each resulting from the product of a Poisson distribution representing the statistical fluctuations of the expected total event yield,
and of Gaussian distributions representing the effect of the systematic uncertainties.
Pseudo-experiments are generated for both hypotheses, taking into account correlations across channels and processes.
The fraction of pseudo-experiments for the signal-plus-background (background-only) hypothesis with LLR
larger than a threshold defines $CL_{s+b}$ ($CL_{b}$).
This threshold is set to the observed (background median) LLR for the observed (expected) limit.
Signal cross-sections for which $CL_s = CL_{s+b}/CL_b < 0.05$ are considered excluded at the 95\%~CL.

\begin{table*}[!htb]
\caption
{
  Expected and observed event yields in the SRI (SRII) signal region, combining the electron and muon channels.
  The expected contribution of resonant (non-resonant) signal corresponding to the
  lowest and highest mass hypotheses considered in this analysis
  and of SM backgrounds are given.
  The first quoted uncertainty gives the uncertainty due to statistics.
  The second one gives the uncertainties due to all other systematic effects,
  symmetrised, regrouped, and summed quadratically, without taking into account possible anticorrelations between systematic uncertainties
  and between processes, for the purpose of this table.
}
\begin{center}
\begin{tabular}{l|*{1}{r@{}c@{ $\pm$ }r@{ $\pm$ }l}|*{1}{r@{}c@{ $\pm$ }r@{ $\pm$ }l}}
\hline\hline
                                                               & \multicolumn{4}{c|}{SRI} & \multicolumn{4}{c}{SRII} \\
\hline\hline
Resonant signal, $m(S)=500$~GeV, $\mfmet=0$~GeV                & $253$ && $5$& $34$            & \multicolumn{4}{c}{--}\\
Resonant signal, $m(S)=500$~GeV, $\mfmet=100$~GeV              & $186$ && $4$& $24$            & \multicolumn{4}{c}{--}\\
\hline
Non-resonant signal, $\mvmet=0$~GeV                            & \multicolumn{4}{c|}{--}       & $2430$ && $130$& $210$\\
Non-resonant signal, $\mvmet=1000$~GeV                         & \multicolumn{4}{c|}{--}       & $8.4$ && $0.1$& $0.8$\\
\hline
\ttbar\                                                        & $190$ && $7$& $40$            & $94$ && $5$& $19$\\
Single-top $s$-channel                                         & \multicolumn{4}{c|}{$<0.05$}  & \multicolumn{4}{c}{$<0.05$}\\
Single-top $t$-channel                                         & \multicolumn{4}{c|}{$<0.10$}  & \multicolumn{4}{c}{$<0.10$}\\
Single-top $Wt$                                                & $19$ && $4$& $14$             & $10$ && $3$& $11$\\
$W$+light-quarks                                               & $2$ && $2$& $4$               & $3$ && $3$& $4$\\
$W$+$bb$                                                       & $10$ && $3$& $5$              & $9$ && $3$& $7$\\
$W$+$cc$                                                       & $5$ && $3$& $3$               & $2$ && $7$& $2$\\
$W$+$c$                                                        & $12$ && $5$& $8$              & $4$ && $2$& $4$\\
Diboson                                                        & $1.3$ && $0.6$& $0.7$         & $1.0$ && $0.5$& $0.5$\\
$Z$+jets                                                       & \multicolumn{4}{c|}{$<4$}     & \multicolumn{4}{c}{$<4$}\\
Multijet                                                       & \multicolumn{4}{c|}{$<0.6$}   & \multicolumn{4}{c}{$<1.3$}\\
\hline
Total background                                               & $240$ && $10$ & $50$          & $124$ && $11$ & $27$\\
\hline
Data                                                           & $238$ & \multicolumn{3}{c|}{} & $133$ & \multicolumn{3}{c}{}\\
\hline\hline
\end{tabular}
\end{center}
\label{tab:Yields}
\end{table*}

\begin{figure*}[!htb]
\centering
\includegraphics[width=0.44\textwidth]{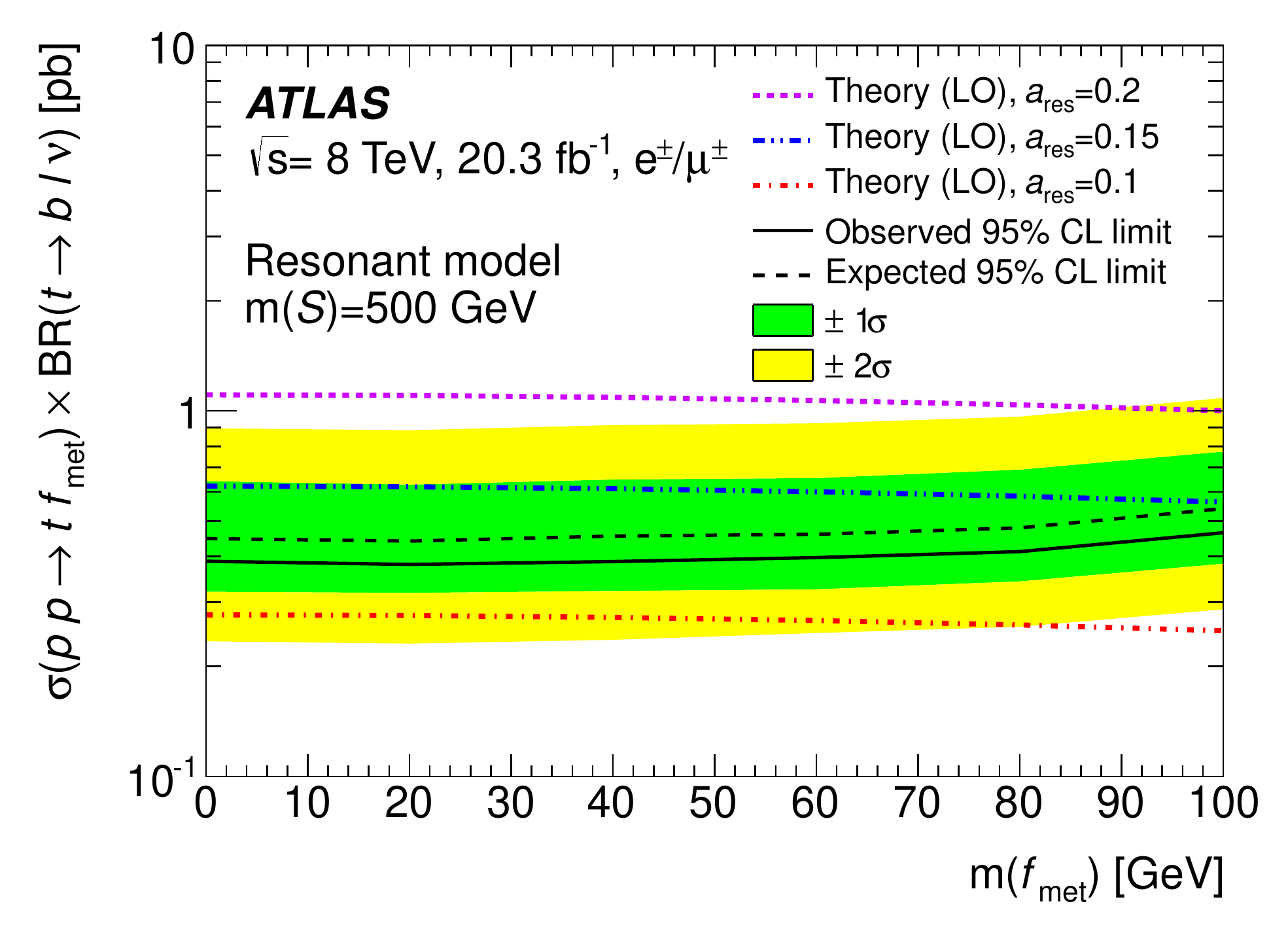}
\includegraphics[width=0.44\textwidth]{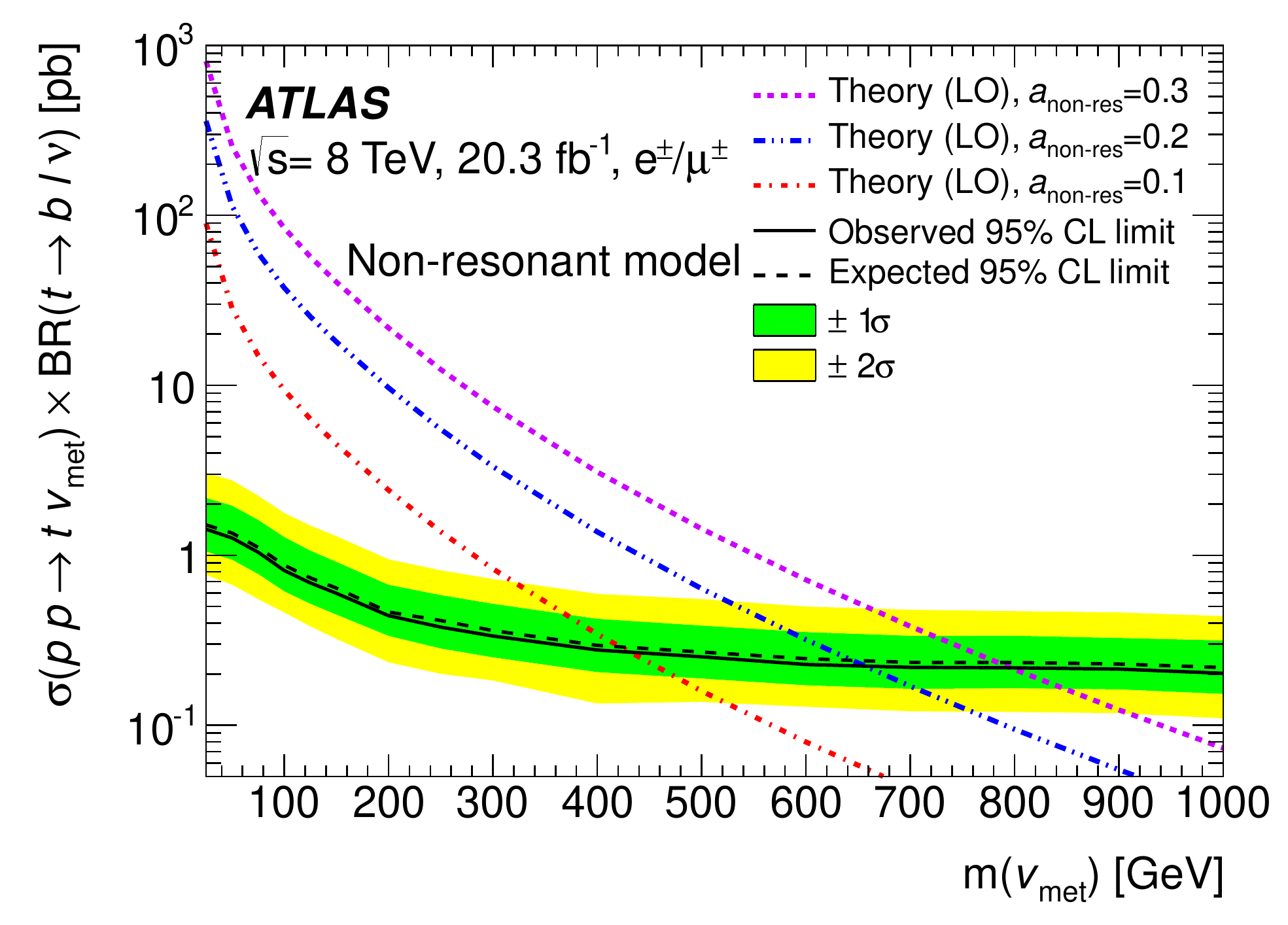}
\caption
{
  Observed and expected limits on the cross section times branching ratio (left) for the resonant
  model with $m(S)=500$~GeV and (right) for the non-resonant model, as a function of the mass of $\fmet$ and $\vmet$, respectively.
  The predicted signal cross-sections for different coupling strengths are also shown.
}
\label{fig:XsecLimit}
\end{figure*}

Figure~\ref{fig:XsecLimit} shows the expected and observed 95\%~CL excluded cross-section times branching ratio
as a function of the mass of the invisible state,
for each of the two signal models.
In the case of the resonant model, cross-sections corresponding to an effective coupling strength $\ares=0.2$ are excluded
in the whole mass range, but not cross-sections corresponding to $\ares=0.1$.
For the non-resonant model, cross-sections corresponding to $\anonres=0.1$ ($0.2$, $0.3$)
are excluded up to $\mvmet=432$~GeV ($657$~GeV, $796$~GeV).

\begin{figure*}[!htb]
\centering
\includegraphics[width=0.44\textwidth]{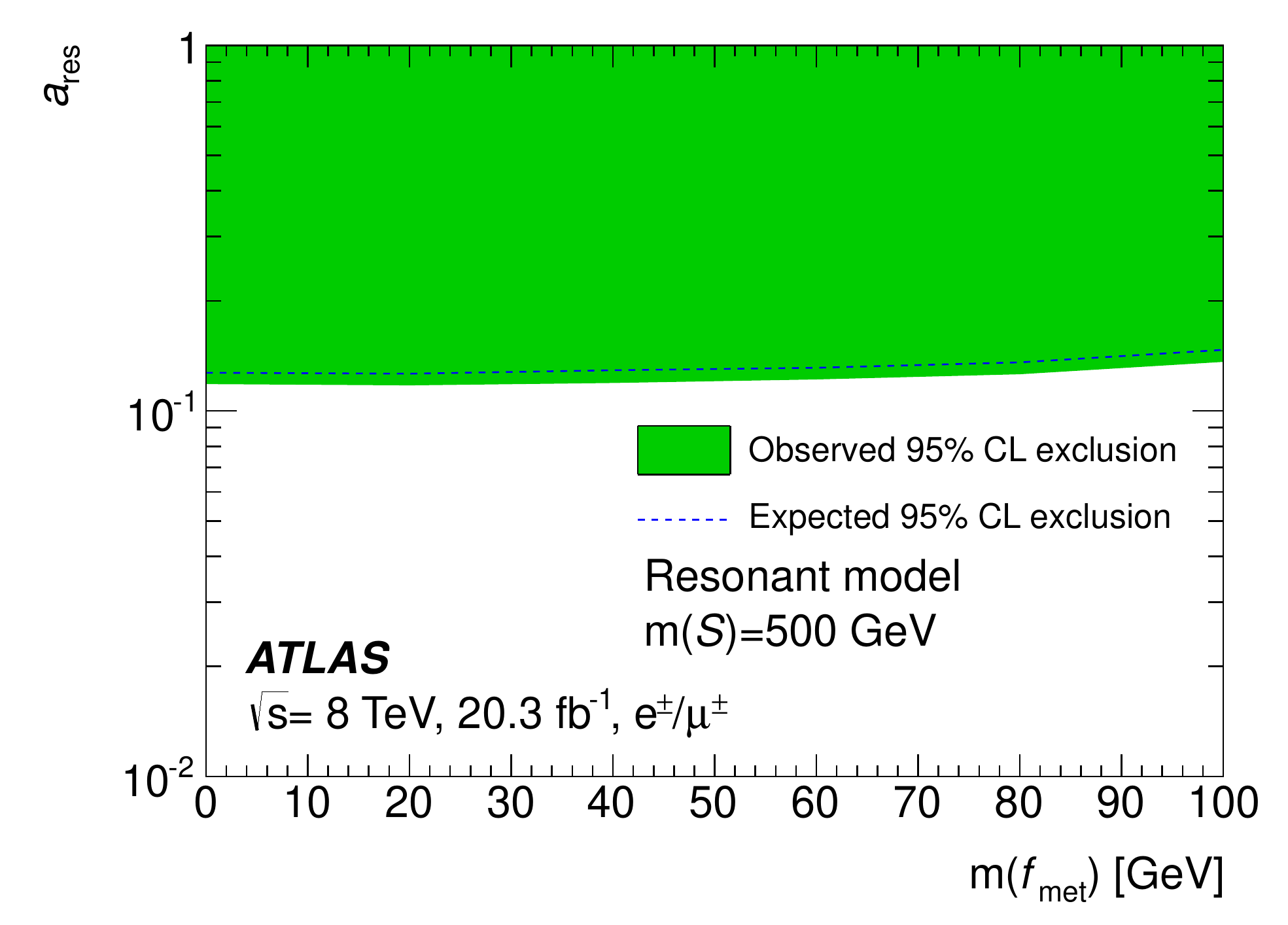}
\includegraphics[width=0.44\textwidth]{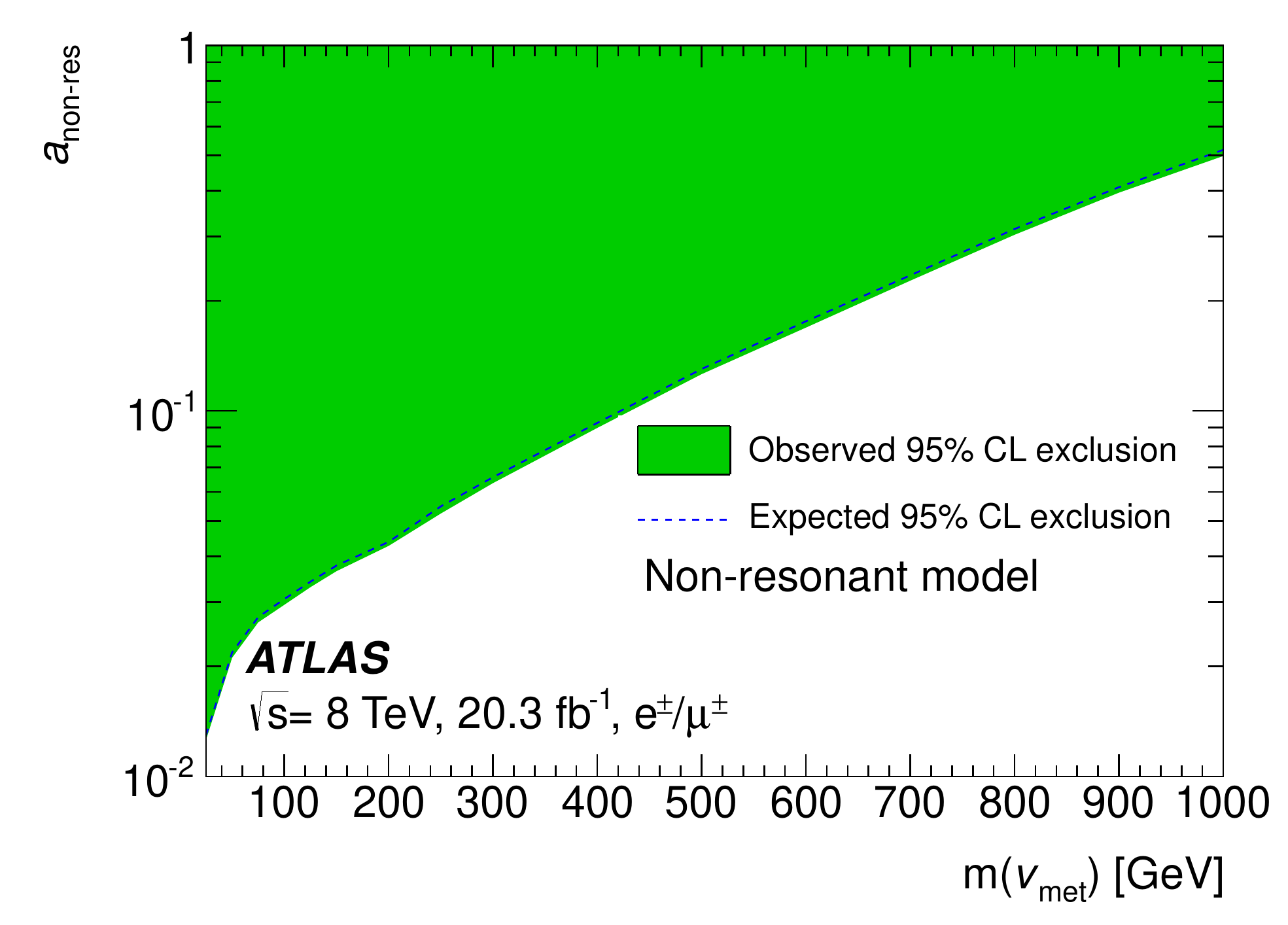}
\caption
{
  Observed and expected excluded coupling strengths (left) for the resonant model with $m(S)=500$~GeV
  and (right) for the non-resonant model, as a function of the mass of $\fmet$ and $\vmet$, respectively.
}
\label{fig:CoupLimit}
\end{figure*}

The cross-sections are proportional to the square of the effective coupling.  
Thus, a 95\%~CL upper limit on $\ares$ and $\anonres$ as a function of the mass of the invisible states is extracted.
The results are shown in Fig.~\ref{fig:CoupLimit}.
This upper limit is set assuming that the coupling has no effect on the signal acceptance modelling.
In the case of the resonant model, in which the increase of the resonance width with increasing coupling strength changes the signal kinematics,
this assumption is validated by using two dedicated simulated samples produced with $\ares=0.5$ and $\ares=1.0$ instead of $\ares=0.2$.
These two hypotheses are excluded at 95\%~CL with the same limit-setting procedure.
Since the kinematic distributions are similar in the whole $\mfmet$ range, this assumption is valid for all values of the $\fmet$ mass.
Tables~\ref{tab:S1R_CouplingExcl} and \ref{tab:S4R_CouplingExcl} give the expected and observed 95\%~CL upper limits on the effective coupling
as a function of the mass of the invisible state, for the resonant and non-resonant model, respectively.

\clearpage
\newpage
\begin{table*}[p]
\caption
{
Expected and observed 95\%~CL limits on the effective coupling $\ares$ as a function of the mass of the invisible state
for the resonant model.
}
\begin{center}
\begin{tabular}{p{0.15\textwidth}*{2}{|p{0.1\textwidth}}}
\hline\hline
\multirow{2}{*}{$\mfmet$~[GeV]} & \multicolumn{2}{r}{95\%~CL upper limit on $\ares$}   \\
& Expected  & Observed   \\
\hline\hline
0                   & 0.13                   & 0.12 \\
20                  & 0.13                   & 0.12 \\
40                  & 0.13                   & 0.12 \\
60                  & 0.13                   & 0.12 \\
80                  & 0.14                   & 0.13 \\
100                 & 0.15                   & 0.14 \\
\hline\hline
\end{tabular}
\end{center}
\label{tab:S1R_CouplingExcl}
\end{table*}

\begin{table*}[p]
\caption
{
  Expected and observed 95\%~CL limits on the effective coupling $\anonres$ as a function of the mass of the invisible state
  for the non-resonant model.
}
\begin{center}
\begin{tabular}{p{0.15\textwidth}*{2}{|p{0.1\textwidth}}}
\hline\hline
\multirow{2}{*}{$\mvmet$~[GeV]} & \multicolumn{2}{r}{95\%~CL upper limit on $\anonres$}   \\
& Expected  & Observed   \\
\hline\hline
0                   & 0.03                  & 0.029 \\
25                  & 0.013                  & 0.013 \\
50                  & 0.022                  & 0.021 \\
75                  & 0.027                  & 0.026 \\
100                 & 0.031                  & 0.030 \\
125                 & 0.034                  & 0.033 \\
150                 & 0.038                  & 0.036 \\
200                 & 0.044                  & 0.043 \\
250                 & 0.055                  & 0.052 \\
300                 & 0.066                  & 0.063 \\
400                 & 0.093                  & 0.090 \\
500                 & 0.13                   & 0.13  \\
600                 & 0.18                   & 0.17  \\
700                 & 0.24                   & 0.23  \\
800                 & 0.32                   & 0.30  \\
900                 & 0.41                   & 0.40  \\
1000                & 0.52                   & 0.50  \\
\hline\hline
\end{tabular}
\end{center}
\label{tab:S4R_CouplingExcl}
\end{table*}

\clearpage
\section{Summary and conclusion}
\label{sec:conclusions}

Monotop events are searched for in the $\sqrt{s}=\mathrm{8~TeV}$ $pp$ collision data collected in 2012 by the ATLAS experiment at the LHC corresponding
to an integrated luminosity of 20.3~$\mathrm{fb}^{-1}$.
Two classes of signal models are studied, producing right-handed top quarks together with exotic neutral particles
giving rise to missing energy.
The semi-leptonic decay mode of the top quark is exploited:
events with one isolated electron or muon and one $b$-tagged jet are selected.
No significant deviation from the Standard Model predictions is observed.
Upper limits on the signal cross-sections and on the corresponding effective couplings are set at 95\% CL using the $CL_s$ method.
In the case of the production of a 500~GeV spin-0 resonance, effective coupling strengths above $\ares=0.15$ are excluded
for a mass of the invisible spin-1/2 state between 0~GeV  and 100~GeV.
In the case of non-resonant production, effective coupling strengths above $\anonres=0.1$, 0.2, and 0.3
are excluded for a mass of the invisible spin-1 state up to 432~GeV, 657~GeV, and 796~GeV, respectively.
The observed 95\% CL limits are compatible with the expectations.
%

\phantomsection 
\addcontentsline{toc}{section}{Acknowledgements} 
\section*{Acknowledgements}

We thank CERN for the very successful operation of the LHC, as well as the
support staff from our institutions without whom ATLAS could not be
operated efficiently.

We acknowledge the support of ANPCyT, Argentina; YerPhI, Armenia; ARC,
Australia; BMWFW and FWF, Austria; ANAS, Azerbaijan; SSTC, Belarus; CNPq and FAPESP,
Brazil; NSERC, NRC and CFI, Canada; CERN; CONICYT, Chile; CAS, MOST and NSFC,
China; COLCIENCIAS, Colombia; MSMT CR, MPO CR and VSC CR, Czech Republic;
DNRF, DNSRC and Lundbeck Foundation, Denmark; EPLANET, ERC and NSRF, European Union;
IN2P3-CNRS, CEA-DSM/IRFU, France; GNSF, Georgia; BMBF, DFG, HGF, MPG and AvH
Foundation, Germany; GSRT and NSRF, Greece; ISF, MINERVA, GIF, I-CORE and Benoziyo Center,
Israel; INFN, Italy; MEXT and JSPS, Japan; CNRST, Morocco; FOM and NWO,
Netherlands; BRF and RCN, Norway; MNiSW and NCN, Poland; GRICES and FCT, Portugal; MNE/IFA, Romania; MES of Russia and ROSATOM, Russian Federation; JINR; MSTD,
Serbia; MSSR, Slovakia; ARRS and MIZ\v{S}, Slovenia; DST/NRF, South Africa;
MINECO, Spain; SRC and Wallenberg Foundation, Sweden; SER, SNSF and Cantons of
Bern and Geneva, Switzerland; NSC, Taiwan; TAEK, Turkey; STFC, the Royal
Society and Leverhulme Trust, United Kingdom; DOE and NSF, United States of
America.

The crucial computing support from all WLCG partners is acknowledged
gratefully, in particular from CERN and the ATLAS Tier-1 facilities at
TRIUMF (Canada), NDGF (Denmark, Norway, Sweden), CC-IN2P3 (France),
KIT/GridKA (Germany), INFN-CNAF (Italy), NL-T1 (Netherlands), PIC (Spain),
ASGC (Taiwan), RAL (UK) and BNL (USA) and in the Tier-2 facilities
worldwide.

\phantomsection 
\addcontentsline{toc}{section}{References} 
\bibliographystyle{atlas4epjc}
\bibliography{Bibliography}

\clearpage
\onecolumn
\begin{flushleft}
{\Large The ATLAS Collaboration}

\bigskip

G.~Aad$^{\rm 85}$,
B.~Abbott$^{\rm 113}$,
J.~Abdallah$^{\rm 152}$,
S.~Abdel~Khalek$^{\rm 117}$,
O.~Abdinov$^{\rm 11}$,
R.~Aben$^{\rm 107}$,
B.~Abi$^{\rm 114}$,
M.~Abolins$^{\rm 90}$,
O.S.~AbouZeid$^{\rm 159}$,
H.~Abramowicz$^{\rm 154}$
% ...
\bigskip
\\
% ...
$^{11}$ Institute of Physics, Azerbaijan Academy of Sciences, Baku, Azerbaijan\\
% ...
$^{85}$ CPPM, Aix-Marseille Universit{\'e} and CNRS/IN2P3, Marseille, France\\
% ...
$^{90}$ Department of Physics and Astronomy, Michigan State University, East Lansing MI, United States of America\\
% ...
$^{107}$ Nikhef National Institute for Subatomic Physics and University of Amsterdam, Amsterdam, Netherlands\\
% ...
$^{113}$ Homer L. Dodge Department of Physics and Astronomy, University of Oklahoma, Norman OK, United States of America\\
$^{114}$ Department of Physics, Oklahoma State University, Stillwater OK, United States of America\\
% ...
$^{117}$ LAL, Universit{\'e} Paris-Sud and CNRS/IN2P3, Orsay, France\\
% ...
$^{152}$ Institute of Physics, Academia Sinica, Taipei, Taiwan\\
% ...
$^{154}$ Raymond and Beverly Sackler School of Physics and Astronomy, Tel Aviv University, Tel Aviv, Israel\\
% ...
$^{159}$ Department of Physics, University of Toronto, Toronto ON, Canada\\
% ...
\end{flushleft}


\end{document}
